\documentstyle[preprint,aps]{revtex}

\begin{document}

\draft


\tighten

\title{Constraint from Lamb Shift and Anomalous Magnetic Moment
 on Radiatively Induced Lorentz and 
$CPT$ Violation in Quantum Electrodynamics }

\author{ W.F. Chen\renewcommand{\thefootnote}{\dagger}\footnote{E-mail: 
wchen@theory.uwinnipeg.ca} and 
G. Kunstatter\renewcommand{\thefootnote}{\ddagger}\footnote{E-mail: 
gabor@theory.uwinnipeg.ca}}
 
\address{ Department of Physics, University of Winnipeg, Winnipeg, 
Manitoba, Canada R3B 2E9 \\
   and\\
Winnipeg Institute for Theoretical Physics, Winnipeg, Manitoba}

\maketitle

\begin{abstract}
We investigate the precisely measured anomalous magnetic moment
and Lamb shift as tests for the possible existence of the radiatively 
induced Lorentz and $CPT$ violation effects in quantum electrodynamics.
To this end we calculate the one-loop vacuum polarization tensor and 
vertex radiative correction in dimensional reduction and on-shell 
renormalization scheme. We explicitly show how the Lorentz and 
$CPT$ violation sector affects the anomalous magnetic moment and Lamb shift.  
Remarkably, we find  infrared divergences coming from 
Lorentz and $CPT$ violating term that do not cancel in physical cross 
sections. This result appears to place stringent constraints on the type
of Lorentz/CPT violating terms that can be added to the QED Lagrangian.
\vspace{3mm}

\noindent PACS number(s): 11.10, 12.20, 11.15.B, 13.40.k
  
\end{abstract}

\vspace{3ex}

\section{Introduction}

Special relativity is one of the most important cornerstones of 
modern physics. Its algebraic foundation, Lorentz transformation 
invariance and discrete $CPT$ symmetry, has become a fundamental 
and indispensable axiom in constructing  relativistic quantum 
field theories.  However, physics is a science based on experimental 
observation. All of its principles must be tested and can only be 
confirmed to the accuracy of the experimental data. With the 
availability of increasingly accurate experimental data and the 
discovery of new phenomena, it is conceivable that even the most
fundamental  principles may someday have to be modified or even
abandoned.  It is partly in this spirit that  there has recently
been increasing interest in the possible breaking of Lorentz
symmetry. As emphasized by Jackiw\cite{ref1}, 
the availability of higher precision
instruments makes it possible to carry out a more precise test on the
principle of special relativity, hence it is not unreasonable to make
a theoretical investigation of the possible violation of Lorentz and $CPT$
symmetry. In fact, if the Standard Model is considered as the low-energy
limit of a more fundamental theory constructed from strings, the spontaneous
breaking of Lorentz symmetry can occur\cite{ref2}. 

A theoretical framework, the extension of the Standard Model
with Lorentz and $CPT$ breaking term has already 
been constructed\cite{ref3}. Based on this
model a series of predictions for  possible signals of $CPT$ and Lorentz
violation have been suggested, including neutral-meson 
oscillations\cite{ref4},
clock-comparison experiments\cite{ref5}, hydrogen and anti-hydrogen 
spectroscopy\cite{ref6} and Penning trap experiments\cite{ref7} etc.  
The QED sector of this extended Standard Model contains a Lorentz and $CPT$
violating Chern-Simons term 
\begin{equation}
{\cal L}_{CS}
=1/2 k_\mu\epsilon^{\mu\nu\rho\lambda}F_{\nu\rho}A_{\lambda}
\label{eq: cs1}
\end{equation}
  which
can lead to the birefringence of light in vacuum and was introduced
earlier in QED\cite{ref8}. However, this Chern-Simons term gives a negative
contribution to the conserved energy and hence makes the theory
instable. Thus the coefficient $k_\mu$ should set to zero. The experimental
searches for cosmological birefringence place a very stringent
limit on $k_\mu$. 

Even if the Chern-Simons  term  Eq.\,(\ref{eq: cs1}) vanishes at classical 
level, it can be generated from the radiative corrections due to an explicit
Lorentz and $CPT$ violation 
term $b^\mu\bar{\psi}\gamma_\mu\gamma_5\psi$ in the 
fermion sector. A series of calculations of the generation of this term from
 radiative corrections was carried out, and appeared to give 
regularization dependent results\cite{ref9,ref10,ref11,ref12,ref13}. 
Moreover, the experimental upper limit on $b_\mu$ is far less 
stringent than on $k_\mu$. Of course, 
in an extended Standard Model, such radiative
corrections from every species of quarks and leptons must cancel if the 
the theory is anomaly-free. However, when QED is considered not embedded
in a large gauge theory, it is necessary to investigate the upper limit
of $b_\mu$ using the experiment data within QED itself. Significantly,
explicit calculations have shown \cite{ref9} that the 
radiatively induced Chern-Simons term in QED escapes 
the no-go theorem\cite{ref14} which 
prohibits the generation of Lorentz and $CPT$ violation term for any gauge
invariant $CPT$-odd interaction. 

It is well known  that two of the most remarkable accomplishments of QED are
the explanations for the anomalous magnetic moment of electron and the Lamb
shift, both of which have been measured with highly precise accuracy. 
The introduction
of a gauge invariant $CPT$-odd term, $b_\mu\bar{\psi}\gamma^{\mu}\gamma_5\psi$,
in the fermionic part would inevitably affect the theoretical values of
the anomalous magnetic moment and of the Lamb shift. The purpose of 
this paper is therefore to propose the use of the experimental data for
the anomalous magnetic moment and the Lamb shift to directly impose an
upper limit on $b_\mu$ and indirectly on $k_\mu$. We think this is quite
significant since it provides another experimental test on the possible
existence of radiatively induced Lorentz and $CPT$ violation effects.

The classical action of QED with the inclusion of $CPT$-odd term in the 
fermionic sector is\cite{ref9,ref12} 
\begin{eqnarray}
{\cal L}=-\frac{1}{4}F_{\mu\nu}F^{\mu\nu}+\bar{\psi}
\left(i\partial\hspace{-2.3mm}/-m-
b\hspace{-2mm}/\gamma_5\right)\psi-e\bar{\psi}A\hspace{-2.3mm}/\psi,
\end{eqnarray}
where $b_{\mu}$ is a constant four vector with a fixed orientation 
in space-time.
The term $\bar{\psi}b\hspace{-1mm}\gamma_5\psi$ is gauge invariant, 
but it explicitly
violates both Lorentz and $CPT$ symmetries, since $b_{\mu}$ picks 
up a preferred direction in space-time.
The theory is explicitly gauge invariant under the usual gauge transformation
\begin{eqnarray}
\psi (x)\longrightarrow e^{-ie\Lambda (x)}\psi (x), ~~
\bar{\psi} (x)\longrightarrow e^{ie\Lambda (x)}\bar{\psi} (x), ~~
A_{\mu}(x)\longrightarrow A_{\mu}(x)+\partial_\mu \Lambda (x).
\end{eqnarray}
The gauge symmetry determines that the vacuum polarization tensor must
take the following form,
\begin{eqnarray}
\Pi_{\mu\nu}(p,b)&=&\epsilon_{\mu\nu\lambda\rho}p^{\lambda}b^{\rho}A(p,b)
+\left(p^2g_{\mu\nu}-p_{\mu}p_{\nu}\right)B(p,b)\nonumber\\
&& +\left(p_\mu b_\nu+p_\nu b_\mu-
\frac{p{\cdot}b}{p^2}p_{\mu}p_\nu
-\frac{p^2}{p{\cdot}b}b_\mu b_\nu\right)C(p,b).
\label{eq3}
\end{eqnarray}
The second term in the r.h.s of the above equation screens
the electric charge. In standard QED, this
term will lead to the Uehling potential correction to the 
Coulomb interaction at low-energy,
and hence give a partial contribution  to the Lamb shift 
when this quantum correction is applied to  a
Hydrogen-like atom.

For the vertex radiative correction, 
the introduction of the term $b^\mu\bar{\psi}\gamma_\mu\gamma_5\psi$ and the
bosonic symmetry $p\longleftrightarrow q$, imply that
the quantum on-shell vertex should have the following tensor structure:
\begin{eqnarray}
\bar{u}(q)\Lambda_\mu u(p)&=&\bar{u}(q)
\left[F_1^{\prime}\gamma_\mu+F_2^{\prime}
\frac{p_\mu+q_\mu}{m}+F_3^{\prime} \frac{b_\mu}{m}
+F_4^{\prime}\gamma_5\frac{b_\mu}{m}\right.\nonumber\\ 
&&\left.+F_5^{\prime}
\gamma_\mu\gamma_5+F_6^{\prime}\sigma_{\mu\nu}\frac{b^{\nu}}{m}
+F_7^{\prime}\sigma_{\mu\nu}\gamma_5\frac{b^{\nu}}{m}\right]u(p)\nonumber\\
&=&\bar{u}(q)\left[F_1\gamma_\mu+F_2i\sigma_{\mu\nu}l^\nu +F_3\frac{b_\mu}{m}
+F_4\gamma_5\frac{b_\mu}{m} \right.\nonumber\\
&&\left.+F_5\gamma_\mu\gamma_5+F_6\sigma_{\mu\nu}
\frac{b^{\nu}}{m}
+F_7i\epsilon_{\mu\nu\lambda\rho}\sigma^{\lambda\rho}
\frac{b^{\nu}}{m}\gamma_5\right]u(p),
\label{eq4}
\end{eqnarray}
where $F_i^{\prime}$ and $F_i$, $i=1,\cdots,7$ are the scalar functions of 
$m$, $b^2$, $l^2$, $l\cdot b$ and $(p+q){\cdot}b$ 
with $l_\mu{\equiv}q_\mu-p_\mu$, 
$\sigma_{\mu\nu}=i/2[\gamma_\mu,\gamma_\nu]$. Due to the existence
of the constant vector $b_\mu$, the form factors $F_i$ do not
depend only on the the Lorentz invariant $l^2$. 
Note that in Eq.\,(\ref{eq4}) we have made use of the following 
Gordon identities,
\begin{eqnarray}
\bar{u}(q)\gamma_\mu u(p)&=&\frac{1}{2m}\bar{u}(q)\left[(p+q)_\mu+
i\sigma_{\mu\nu}l^{\nu}\right]u(p);
\label{eqgor1}\\
\bar{u}(q)\gamma_\mu \gamma_5 u(p)&=&\frac{1}{2m}\bar{u}(q)\left[l_\mu
\gamma_5+i\sigma_{\mu\nu}(p+q)^{\nu}\gamma_5\right]u(p),
\label{eqgor2}
\end{eqnarray}
and the self-dual relation $\sigma_{\mu\nu}\gamma^5
=i/2\epsilon_{\mu\nu\lambda\rho}\sigma^{\lambda\rho}$.
The $F_2$ term of Eq.\,(\ref{eq4}) will lead to the famous
Schwinger anomalous magnetic moment term. Furthermore, as shown in Fig.\,1, the
combination of radiative corrections of vertex and the correction to the 
electromagnetic coupling from the vacuum polarization tensor
will yield the Lamb shift.

The paper is organized as follows: 
In Sect.\,II we calculate the one-loop vacuum polarization tensor to 
second order in $b_\mu$. Some of the calculation techniques are illustrated 
and the screening effects on the electric charge of vacuum polarization 
tensor are discussed. Sect.\,III is devoted to the somewhat lengthy calculation
of the one-loop quantum correction to the on-shell vertex to second order
in $b_\mu$. As done in standard QED, the radiative correction is extracted 
using the mass-shell renormalization scheme. In preparation for discussing 
the anomalous magnetic moment and the Lamb shift, we give the expansion of 
the on-shell vertex radiative correction at small photon momentum. Sects.\,IV 
and V contain detailed discussions on the effects of $CPT$-odd term on the 
anomalous magnetic moment and Lamb shift. The explicit $b$-dependence of the 
anomalous magnetic moment and the Lamb shift is presented.
As in conventional QED, the results contain infrared (IR) divergences,
so in Sect.\,VI we consider the soft photon emission of bremsstrahlung
and demonstrate that, contrary to what happens in conventional QED, the IR 
divergences contained in the
$b$-dependent part do not appear to cancel. Finally, we summarize the result,
discuss the other possible effects induced by the $CPT$-odd term
and emphasize the constraints from the anomalous magnetic moment
and the Lamb shift on the existence of Lorentz and $CPT$ violation
in the electromagnetic interaction  due to the non-cancellation
of IR divergence.

\section{vacuum polarization tensor}

The one-loop vacuum polarization tensor is read as
\begin{eqnarray}
\Pi_{\mu\nu}(p,b)=e^2\mu^{2\epsilon_{\rm UV}}
\int \frac{d^n k}{(2\pi)^n}\mbox{Tr}\left[
\gamma_\mu S_b(k)\gamma_\nu S_b(k+p)\right],
\end{eqnarray} 
where we have chosen dimensional regularization, 
$\epsilon_{\rm UV}{\equiv}2-n/2$; $S_b(k)$ is the $b_{\mu}$-exact fermionic 
propagator utilized by Jackiw and Kosteleck\'{y}
in calculating the radiatively induced Lorentz and $CPT$ violating 
Chern-Simons term\cite{ref9},
\begin{eqnarray}
S_b(k)=\frac{i}{k\hspace{-2.3mm}/-m-b\hspace{-2mm}/\gamma_5}.
\end{eqnarray} 
However, this propagator will give rise to complications
when performing Wick rotation in order to
 evaluate the Feynman integral\cite{ref12}. 
Thus motivated by the fact that the magnitude of  $b_{\mu}$ should be
small (compared to $m$)
as well as the fact   the vacuum polarization tensor 
should be  quadratic in $b$ to leading order, we make use of the following
identity for the operators (or matrices) $A$ and $B$,
\begin{eqnarray}
\frac{1}{A+B}=\frac{1}{A}-\frac{1}{A}B\frac{1}{A+B}=
\frac{1}{A}-\frac{1}{A}B\frac{1}{A}+\frac{1}{A}B\frac{1}{A}B\frac{1}{A+B}
={\cdots} \, .
\label{eq6}
\end{eqnarray} 
 The vacuum polarization tensor 
up to  second order $b$ can be  written as
\begin{eqnarray}
\Pi_{\mu\nu}(p,b)&=&
e^2\mu^{2\epsilon_{\rm UV}}\int \frac{d^n k}{(2\pi)^n}\mbox{Tr}\left[
\gamma_\mu \left(\frac{1}{k\hspace{-2.3mm}/-m}
+\frac{1}{k\hspace{-2.3mm}/-m}b\hspace{-2.3mm}/
\gamma_5\frac{1}{k\hspace{-2.3mm}/-m}\right.\right.\nonumber\\
&&\left.+\frac{1}{k\hspace{-2.3mm}/-m}b\hspace{-2mm}/\gamma_5
\frac{1}{k\hspace{-2.3mm}/-m}
b\hspace{-2mm}/\gamma_5 \frac{1}{k\hspace{-2.3mm}/-m}\right)
\gamma_{\nu}
\left(\frac{1}{k\hspace{-2.3mm}/+p\hspace{-2mm}/-m}
+\frac{1}{k\hspace{-2.3mm}/+p\hspace{-2mm}/ -m}
b\hspace{-2mm}/\gamma_5
\frac{1}{k\hspace{-2.3mm}/+p\hspace{-2mm}/ -m}\right.\nonumber\\
&&\left.\left.
+\frac{1}{k\hspace{-2.3mm}/+p\hspace{-2mm}/ -m}b\hspace{-2mm}/
\gamma_5 \frac{1}{k\hspace{-2.3mm}/+p\hspace{-2mm}/ -m}
b\hspace{-2mm}/\gamma_5 \frac{1}{k\hspace{-2.3mm}/+p\hspace{-2mm}/-m}
\right)\right]\nonumber\\
&{\equiv}&\Pi^{(0)}_{\mu\nu}(p)+\Pi^{(1)}_{\mu\nu}(p,b)
+\Pi^{(2)}_{\mu\nu}(p,b),
\end{eqnarray}
Where the Feynman diagrams corresponding 
to $\Pi_{\mu\nu}^{(i)}$, $i=0,1,2$
is illustrated in Fig.\,2. 
$\Pi^{(0)}_{\mu\nu}(p)$ is just the vacuum polarization tensor in
the conventional QED,
\begin{eqnarray}
\Pi^{(0)}_{\mu\nu}(p)&=&e^2\mu^{2\epsilon_{\rm UV}}\int \frac{d^n k}{(2\pi)^n}
\mbox{Tr}\left[\gamma_{\mu}\frac{1}{k\hspace{-2.3mm}/-m}
\gamma_{\nu}\frac{1}{k\hspace{-2.3mm}/+p\hspace{-2.3mm}/ -m}\right]\nonumber\\
&=&4 e^2\mu^{2\epsilon_{\rm UV}}
 \int \frac{d^n k}{(2\pi)^n}\frac{2k_\mu k_\nu+k_\mu p_\nu+k_\nu p_\mu
-g_{\mu\nu}k{\cdot}(k+p)+m^2g_{\mu\nu}}{(k^2-m^2)[(k+p)^2-m^2]}\nonumber\\
&=&\left(p^2g_{\mu\nu}-p_{\mu}p_{\nu}\right)\frac{ie^2}{2\pi^2}
\left(\frac{4\pi\mu^2}{m^2}\right)^{\epsilon_{\rm UV}}
\Gamma(\epsilon_{\rm UV})\int_0^1 d x
\frac{x (1-x)}{\left[1-p^2/m^2 x (1-x)\right]^{\epsilon}}
\end{eqnarray} 
In the four-dimensional limit, we have in the case of $p^2<4m^2$,
\begin{eqnarray}
\Pi^{(0)}_{\mu\nu}(p)&=&\left(p^2g_{\mu\nu}-p_{\mu}p_{\nu}\right)
\frac{ie^2}{2\pi^2}\left[\frac{1}{6}\left(\frac{1}{\epsilon_{\rm UV}}
-\gamma+\ln 4\pi\right)+\frac{1}{6}\ln\frac{\mu^2}{m^2}+\frac{5}{18}+
\frac{2}{3}\frac{m^2}{p^2}\right.\nonumber\\
&&\left.-\left(\frac{8}{3}\frac{m^4}{p^4}
+\frac{2}{3}\frac{m^2}{p^2}-\frac{1}{3}\right)\frac{p/m}{\sqrt{4-p^2/m^2}}
\arctan \frac{p/m}{\sqrt{4-p^2/m^2}}\right].
\label{eq9}
\end{eqnarray}
This is the well-known textbook result\cite{ref15}.

$\Pi^{(1)}_{\mu\nu}(p,b)$ contains the linear terms in $b_\mu$,
\begin{eqnarray}
\Pi^{(1)}_{\mu\nu}(p,b)&=& e^2\mu^{2\epsilon_{\rm UV}}\int\frac{d^nk}{(2\pi)^n}
\left[\mbox{Tr}\left(\gamma_\mu\frac{1}{k\hspace{-2.3mm}/-m}
\gamma_{\nu}\frac{1}{k\hspace{-2.3mm}/+p\hspace{-2mm}/ -m}
b\hspace{-2mm}/\gamma_5\frac{1}{k\hspace{-2.3mm}/+p\hspace{-2mm}/ -m}
\right)\right.\nonumber\\
&&\left.+\mbox{Tr}\left(\gamma_\mu \frac{1}{k\hspace{-2.3mm}/-m}
b\hspace{-2mm}/\gamma_5\frac{1}{k\hspace{-2.3mm}/-m}\gamma_\nu
\frac{1}{k\hspace{-2.3mm}/+p\hspace{-2mm}/ -m}\right)\right].
\end{eqnarray}
This term has been widely investigated using various regularization
methods. Despite the existence of a  regularization ambiguity, depending
on the concrete physical renormalization condition, it indeed   
leads to the radiatively induced Chern-Simons term, which violate both 
Lorentz and $CPT$ symmetries. In dimensional regularization the result is
\cite{ref10,ref12}:
\begin{eqnarray}
\Pi^{(1)}_{\mu\nu}(p,b)&=&\frac{e^2}{2\pi^2}
\epsilon_{\mu\nu\alpha\beta}
p^{\beta}b^{\alpha}\left[\frac{2m}{p}\frac{1}{\sqrt{1-p^2/4m^2}}
\arctan\frac{p}{2m}-\frac{1}{4}\right].
\end{eqnarray}
The radiatively induced Chern-Simons term can be defined
at low-energy (or equivalently large-$m$) limit.

In the following we concentrate on the quadratic term in $b$,
\begin{eqnarray}
\Pi^{(2)}_{\mu\nu}(p,b)&=&e^2\mu^{2\epsilon_{\rm UV}}\int\frac{d^nk}{(2\pi)^n}
\left[\mbox{Tr}\left(\gamma_\mu \frac{1}{k\hspace{-2.3mm}/-m}
\gamma_{\nu}\frac{1}{k\hspace{-2.3mm}/+p\hspace{-2mm}/ -m}
b\hspace{-2mm}/\gamma_5\frac{1}{k\hspace{-2.3mm}/+p\hspace{-2mm}/-m}
b\hspace{-2mm}/\gamma_5\frac{1}{k\hspace{-2.3mm}/+p\hspace{-2mm}/-m}
\right)\right.\nonumber\\
&&+\mbox{Tr}\left(\gamma_\mu \frac{1}{k\hspace{-2.3mm}/-m}
b\hspace{-2mm}/\gamma_5 \frac{1}{k\hspace{-2.3mm}/-m}\gamma_\nu
\frac{1}{k\hspace{-2.3mm}/+p\hspace{-2mm}/-m}
b\hspace{-2mm}/\gamma_5\frac{1}{k\hspace{-2.3mm}/+p\hspace{-2mm}/-m}
\right)\nonumber\\
&&+\left.\mbox{Tr}\left(\gamma_\mu \frac{1}{k\hspace{-2.3mm}/-m}
b\hspace{-2mm}/\gamma_5 \frac{1}{k\hspace{-2.3mm}/-m}
b\hspace{-2mm}/\gamma_5\frac{1}{k\hspace{-2.3mm}/-m}
\gamma_\nu \frac{1}{k\hspace{-2.3mm}/+p\hspace{-2mm}/-m}\right)\right].
\label{eq12}
\end{eqnarray}
The trace calculation  in Eq.\,(\ref{eq12}) is a heavy task,  so we first
make use of the following identities
\begin{eqnarray}
b\hspace{-2mm}/\gamma_5\frac{1}{k\hspace{-2.3mm}/-m}
b\hspace{-2mm}/\gamma_5&=&b\hspace{-2mm}/
\left(\frac{1}{k\hspace{-2.3mm}/-m}-\frac{2m}{k^2-m^2}\right)b\hspace{-2mm}/
=b\hspace{-2mm}/\frac{1}{k\hspace{-2.3mm}/-m}b\hspace{-2mm}/
-\frac{2mb^2}{k^2-m^2};\nonumber\\
b\hspace{-2mm}/\gamma_5 \frac{1}{k\hspace{-2.3mm}/-m}\gamma_\nu
\frac{1}{k\hspace{-2.3mm}/+p\hspace{-2mm}/-m}
b\hspace{-2mm}/\gamma_5
&=&b\hspace{-2mm}/\left(\frac{1}{k\hspace{-2.3mm}/-m}-\frac{2m}{k^2-m^2}
\right)\gamma_\nu \left[\frac{1}{k\hspace{-2.3mm}/+p\hspace{-2mm}-m}-
\frac{2m}{(k+p)^2-m^2}\right]b\hspace{-2mm}/,
\label{eq13}
\end{eqnarray}
and the differential operations
\begin{eqnarray}
\frac{\partial}{\partial p_\mu}\frac{1}{k\hspace{-2.3mm}/+p\hspace{-2mm}/ -m}
&=&-\frac{1}{k\hspace{-2.3mm}/+p\hspace{-2mm}/ -m}\gamma_\mu
\frac{1}{k\hspace{-2.3mm}/+p\hspace{-2mm}/ -m};\nonumber\\
\frac{\partial}{\partial p_\mu} \frac{\partial}{\partial p_\nu}
\frac{1}{k\hspace{-2.3mm}/+p\hspace{-2mm}/ -m}&=&
\frac{1}{k\hspace{-2.3mm}/+p\hspace{-2mm}/ -m}\gamma_\mu
\frac{1}{k\hspace{-2.3mm}/+p\hspace{-2mm}/ -m}\gamma_\nu
\frac{1}{k\hspace{-2.3mm}/+p\hspace{-2mm}/ -m}\nonumber\\
&&+
\frac{1}{k\hspace{-2.3mm}/+p\hspace{-2mm}/ -m}\gamma_\nu
\frac{1}{k\hspace{-2.3mm}/+p\hspace{-2mm}/ -m}\gamma_\mu
\frac{1}{k\hspace{-2.3mm}/+p\hspace{-2mm}/ -m}.
\label{eq14}
\end{eqnarray}
Consequently, $\Pi^{(2)}_{\mu\nu}(p,b)$ can be written in the 
following form,
\begin{eqnarray}
\Pi^{(2)}_{\mu\nu}(p,b)&=&e^2\mu^{2\epsilon_{\rm UV}}
\left[\int\frac{d^nk}{(2\pi)^n}
\mbox{Tr}\left(\gamma_\mu \frac{1}{k\hspace{-2.3mm}/-m}
\gamma_{\nu}\frac{1}{k\hspace{-2.3mm}/+p\hspace{-2mm}/ -m}
b\hspace{-2mm}/\frac{1}{k\hspace{-2.3mm}/+p\hspace{-2mm}/-m}
b\hspace{-2mm}/\frac{1}{k\hspace{-2.3mm}/+p\hspace{-2mm}/-m}
\right)\right.\nonumber\\
&&-\int\frac{d^nk}{(2\pi)^n}\frac{2m b^2}{(k+p)^2-m^2}
\mbox{Tr}\left(\gamma_\mu \frac{1}{k\hspace{-2.3mm}/-m}
\gamma_{\nu}\frac{1}{k\hspace{-2.3mm}/+p\hspace{-2mm}/ -m}
\frac{1}{k\hspace{-2.3mm}/+p\hspace{-2mm}/ -m}\right)
\nonumber\\
&&+\int \frac{d^nk}{(2\pi)^n}\mbox{Tr}
\left(\gamma_\mu \frac{1}{k\hspace{-2.3mm}/-m}b\hspace{-2mm}/
\frac{1}{k\hspace{-2.3mm}/-m}\gamma_\nu
\frac{1}{k\hspace{-2.3mm}/+p\hspace{-2mm}/-m}b\hspace{-2mm}/
\frac{1}{k\hspace{-2.3mm}/+p\hspace{-2mm}/-m}\right)\nonumber\\
&&-\int \frac{d^nk}{(2\pi)^n}\frac{2m}{k^2-m^2}
\mbox{Tr}\left(\gamma_\mu \frac{1}{k\hspace{-2.3mm}/-m}b\hspace{-2mm}/
\gamma_\nu\frac{1}{k\hspace{-2.3mm}/+p\hspace{-2mm}/-m}b\hspace{-2mm}/
\frac{1}{k\hspace{-2.3mm}/+p\hspace{-2mm}/-m}\right)\nonumber\\
&&-\int \frac{d^nk}{(2\pi)^n}\frac{2m}{(k+p)^2-m^2}\mbox{Tr}
\left(\gamma_\mu \frac{1}{k\hspace{-2.3mm}/-m}b\hspace{-2mm}/
\frac{1}{k\hspace{-2.3mm}/-m}\gamma_\nu b\hspace{-2mm}/
\frac{1}{k\hspace{-2.3mm}/+p\hspace{-2mm}/-m}\right)\nonumber\\
&&+\int \frac{d^nk}{(2\pi)^n}\frac{4m^2}{(k^2-m^2)[(k+p)^2-m^2]}\mbox{Tr}
\left(\gamma_\mu \frac{1}{k\hspace{-2.3mm}/-m}b\hspace{-2mm}/
\gamma_\nu b\hspace{-2mm}/\frac{1}{k\hspace{-2.3mm}/+p\hspace{-2mm}/-m}
\right)\nonumber\\
&&+\int \frac{d^nk}{(2\pi)^n}\mbox{Tr}\left(\gamma_\mu
\frac{1}{k\hspace{-2.3mm}/-p\hspace{-2mm}/-m} b\hspace{-2mm}/
\frac{1}{k\hspace{-2.3mm}/-p\hspace{-2mm}/-m} b\hspace{-2mm}/
\frac{1}{k\hspace{-2.3mm}/-p\hspace{-2mm}/-m} b\hspace{-2mm}/
\gamma_\nu \frac{1}{k\hspace{-2.3mm}/-m}\right)\nonumber\\
&&\left.-\int \frac{d^nk}{(2\pi)^n}\frac{2mb^2}{k^2-m^2}
\mbox{Tr}\left(\gamma_\mu \frac{1}{k\hspace{-2.3mm}/-m}
\frac{1}{k\hspace{-2.3mm}/-m}\gamma_\nu \frac{1}{k\hspace{-2.3mm}/-m}
\gamma_\nu b\hspace{-2mm}/
\frac{1}{k\hspace{-2.3mm}/+p\hspace{-2mm}/-m}\right)\right]
\nonumber\\
&=&e^2\mu^{2\epsilon_{\rm UV}}
\left[b^{\alpha}b^{\beta}\frac{\partial}{\partial p^{\alpha}}
\frac{\partial}{\partial p^{\beta}}\int\frac{d^nk}{(2\pi)^n}
\mbox{Tr}\left(\gamma_\mu \frac{1}{k\hspace{-2.3mm}/-m}\gamma_\nu
\frac{1}{k\hspace{-2.3mm}/+p\hspace{-2mm}/-m}\right)\right.\nonumber\\
&&-\int\frac{d^nk}{(2\pi)^n}\frac{2mb^2}{(k+p)^2-m^2}
\mbox{Tr}\left(\gamma_\mu \frac{1}{k\hspace{-2.3mm}/-m}
\gamma_{\nu}\frac{1}{k\hspace{-2.3mm}/+p\hspace{-2mm}/ -m}
\frac{1}{k\hspace{-2.3mm}/+p\hspace{-2mm}/ -m}\right)
\nonumber\\
&&-\int \frac{d^nk}{(2\pi)^n}\frac{2m}{k^2-m^2}
\mbox{Tr}\left(\gamma_\mu \frac{1}{k\hspace{-2.3mm}/-m}b\hspace{-2mm}/
\gamma_\nu\frac{1}{k\hspace{-2.3mm}/+p\hspace{-2mm}/-m}b\hspace{-2mm}/
\frac{1}{k\hspace{-2.3mm}/+p\hspace{-2mm}/-m}\right)\nonumber\\
&&-b^{\alpha}b^{\beta}\frac{\partial}{\partial p^{\alpha}}
\frac{\partial}{\partial p^{\beta}}\int\frac{d^nk}{(2\pi)^n}
\mbox{Tr}\left(\gamma_\mu \frac{1}{k\hspace{-2.3mm}/+p\hspace{-2mm}/-m}
\gamma_\nu\frac{1}{k\hspace{-2.3mm}/-m}\right)\nonumber\\
&&+b^{\alpha}\frac{\partial}{\partial p^{\alpha}}\int 
\frac{d^nk}{(2\pi)^n}\frac{2m}{k^2-m^2}\mbox{Tr}\left(\gamma_\mu 
\frac{1}{k\hspace{-2.3mm}/-m}b\hspace{-2mm}/\gamma_\nu 
\frac{1}{k\hspace{-2.3mm}/+p\hspace{-2mm}/-m}\right)\nonumber\\
&&-b^{\alpha}\frac{\partial}{\partial p^{\alpha}}\int 
\frac{d^nk}{(2\pi)^n}\frac{2m}{(k+p)^2-m^2}\mbox{Tr}\left(\gamma_\mu 
\frac{1}{k\hspace{-2.3mm}/-m}\gamma_\nu b\hspace{-2mm}/ 
\frac{1}{k\hspace{-2.3mm}/+p\hspace{-2mm}/-m}\right)\nonumber\\
&&+\left.\int \frac{d^nk}{(2\pi)^n}\frac{4m^2}{(k^2-m^2)[(k+p)^2-m^2]}\mbox{Tr}
\left(\gamma_\mu \frac{1}{k\hspace{-2.3mm}/-m}b\hspace{-2mm}/
\gamma_\nu b\hspace{-2mm}/\frac{1}{k\hspace{-2.3mm}/+p\hspace{-2mm}/-m}
\right)\right]\nonumber\\
&=&e^2b^{\alpha}\frac{\partial}{\partial p^{\alpha}}\int 
\frac{d^4k}{(2\pi)^4}\left[\frac{2m}{k^2-m^2}\mbox{Tr}\left(\gamma_\mu 
\frac{1}{k\hspace{-2.3mm}/-m}b\hspace{-2mm}/\gamma_\nu 
\frac{1}{k\hspace{-2.3mm}/+p\hspace{-2mm}/-m}\right)\right.\nonumber\\
&&-\left.\frac{2m}{(k+p)^2-m^2}\mbox{Tr}\left(\gamma_\mu 
\frac{1}{k\hspace{-2.3mm}/-m}\gamma_\nu b\hspace{-2mm}/ 
\frac{1}{k\hspace{-2.3mm}/+p\hspace{-2mm}/-m}\right)\right]\nonumber\\
&&+e^2
\int\frac{d^4k}{(2\pi)^4}\left[\frac{4m^2}{(k^2-m^2)[(k+p)^2-m^2]}\mbox{Tr}
\left(\gamma_\mu \frac{1}{k\hspace{-2.3mm}/-m}b\hspace{-2mm}/
\gamma_\nu b\hspace{-2mm}/\frac{1}{k\hspace{-2.3mm}/+p\hspace{-2mm}/-m}
\right)\right.\nonumber\\
&& -\frac{2mb^2}{k^2-m^2}\mbox{Tr}\left(\gamma_\mu 
\frac{1}{(k\hspace{-2.3mm}/-m)^2}\gamma_\nu 
\frac{1}{k\hspace{-2.3mm}/+p\hspace{-2mm}/-m}\right)\nonumber\\
&&\left.-\frac{2mb^2}{(k+p)^2-m^2}
\mbox{Tr}\left(\gamma_\mu \frac{1}{k\hspace{-2.3mm}/-m}
\gamma_{\nu}\frac{1}{(k\hspace{-2.3mm}/+p\hspace{-2mm}/ -m)^2}
\right)\right].
\end{eqnarray}
Taking the $\gamma$-matrix trace and performing Feynman parameterization,
 we get
\begin{eqnarray}
\Pi^{(2)}_{\mu\nu}(p,b)&=&e^2 b^{\alpha}\frac{\partial}{\partial p^{\alpha}}
 \int\frac{d^4k}{(2\pi)^4}\left[
\frac{8m^2}{(k^2-m^2)^2[(k+p)^2-m^2]}
\left(2k_\mu b_\nu+p_\mu b_\nu+b_\mu p_\nu-g_{\mu\nu}p{\cdot}b\right)\right.
\nonumber\\
&&-\left.\frac{8m^2}{(k^2-m^2)[(k+p)^2-m^2]^2}
\left(2k_\mu b_\nu+p_\mu b_\nu-b_\mu p_\nu+g_{\mu\nu}p{\cdot}b\right)
\right]\nonumber\\
&&+e^2\int\frac{d^4k}{(2\pi)^4}\left(\frac{16m^2}{(k^2-m^2)^2[(k+p)^2-m^2]^2}
\left\{2b_\nu\left[k_\mu b{\cdot}(k+p)-b_\mu k{\cdot}(k+p)\right.\right.
\right.\nonumber\\
&&\left.\left.+(k+p)_\mu k{\cdot}b
\right]-b^2\left[2k_\mu k_\nu+k_\mu p_\nu+k_\nu p_\mu-g_{\mu\nu}k{\cdot}(k+p)
\right]+ m^2 (2b_\mu b_\nu-g_{\mu\nu}b^2)\right\}\nonumber\\
&&-\frac{8m^2b^2}{(k^2-m^2)[(k+p)^2-m^2]^3}\left\{
2\left[2k_\mu k_\nu +k_\mu p_\nu+k_\nu p_\mu-g_{\mu\nu}k{\cdot}(k+p)\right]
\right.\nonumber\\
&&+\left.[(k+p)^2+m^2]g_{\mu\nu}\right\}\nonumber\\
&&-\frac{8m^2b^2}{(k^2-m^2)^3[(k+p)^2-m^2]}\left\{
2\left[2k_\mu k_\nu +k_\mu p_\nu+k_\nu p_\mu-g_{\mu\nu}k{\cdot}(k+p)\right]
\right.\nonumber\\
&&\left.\left.+(k^2+m^2)g_{\mu\nu}\right\}\right)\nonumber\\
&=&\left(p^2 g_{\mu\nu}-p_\mu p_\nu\right)\int_0^1 dx\left[ 
\frac{x (1-x)}{[m^2-p^2 x (1-x)]^2}
\left(-i\frac{e^2m^2b^2}{\pi^2}\right)\right.
\nonumber\\
&&+\left.\frac{x^2 (1-x)}{[m^2-p^2 x (1-x)]^2}\frac{2im^2}{\pi^2}
\frac{(p\cdot b)^2}{p^2}\right]+\left(p_\mu b_\nu+p_\nu b_\mu
-\frac{p^2}{p\cdot b}b_\mu b_\nu-\frac{p{\cdot}b}{p^2}p_\mu p_\nu\right)
\nonumber\\
&&{\times}\int_0^1 d x \frac{x^2 (1-x)}{[m^2-p^2 x (1-x)]^2}
\left(-\frac{2i m^2 e^2 p{\cdot}b}{\pi^2}\right)\nonumber\\
&=&\left\{\left(p^2 g_{\mu\nu}-p_\mu p_\nu\right)
\left[\frac{b^2}{p^2}-\frac{\left(p{\cdot}b\right)^2}{p^4}\right]
-\left(p_\mu b_\nu+p_\nu b_\mu-\frac{p^2}{p{\cdot}b}b_\mu b_\nu 
-\frac{p{\cdot}b}{p^2}p_\mu p_\nu\right)\frac{p{\cdot}b}{p^2}\right\}
\nonumber\\
&&\times\frac{2ie^2}{\pi^2}\frac{m^2}{4m^2-p^2}
\left[1+\frac{2p}{\sqrt{4m^2-p^2}}\left(1-\frac{2m^2}{p^2}\right)
\arctan\frac{p}{\sqrt{4m^2-p^2}}\right].
\label{eq19}
\end{eqnarray}
Combining the result of Eq.\,(\ref{eq19}) 
with Eq.\,(\ref{eq9}), the vacuum polarization tensor
in the absence of Lorentz and $CPT$ violation term, we obtain
the polarization tensor associated with the quantum correction
of electric charge
\begin{eqnarray}
\Pi_{\mu\nu}(p,b)=i\left(p^2g_{\mu\nu}-p_\mu p_\nu\right)\omega (p,b)
+{\cdots}
\end{eqnarray}
with
\begin{eqnarray}
\omega (p,b)&=&\frac{2\alpha}{\pi}\left\{\left[
\left(\frac{1}{\epsilon_{\rm UV}}
-\gamma+\ln 4\pi\right)+\frac{1}{6}\ln\frac{\mu^2}{m^2}+\frac{5}{18}+
\frac{2}{3}\frac{m^2}{p^2}-\left(\frac{8}{3}\frac{m^4}{p^4}
+\frac{2}{3}\frac{m^2}{p^2}-\frac{1}{3}\right)\right.\right.\nonumber\\
&&{\times}\left.\frac{p/m}{\sqrt{4-p^2/m^2}}
\arctan \frac{p/m}{\sqrt{4-p^2/m^2}}\right]
+\frac{4}{4-p^2/m^2}\left[\frac{b^2}{p^2}-\frac{\left(p{\cdot}b\right)^2}{p^4}
\right]\nonumber\\
&&{\times}\left.\left[1+\left(1-\frac{2m^2}{p^2}\right)
\frac{2p/m}{\sqrt{4-p^2/m^2}}
\arctan\frac{p/m}{\sqrt{4-p^2/m^2}}\right]\right\}.
\end{eqnarray}
The on-shell renormalization condition
\begin{eqnarray}
\omega_R (p,b)|_{p^2=0}=0
\end{eqnarray}
gives the radiative correction part of the vacuum polarization tensor,
\begin{eqnarray}
\omega_R (p,b)&=&\frac{2\alpha}{\pi}\left\{\frac{5}{18}+
\frac{2}{3}\frac{m^2}{p^2}-\left(\frac{8}{3}\frac{m^4}{p^4}
+\frac{2}{3}\frac{m^2}{p^2}-\frac{1}{3}\right)\frac{p/m}{\sqrt{4-p^2/m^2}}
\arctan \frac{p/m}{\sqrt{4-p^2/m^2}}\right.\nonumber\\
&&+\frac{4}{4-p^2/m^2}\left[\frac{b^2}{p^2}
-\frac{\left(p{\cdot}b\right)^2}{p^4}
\right]\left[1-\frac{p^2}{3m^2}+\left(1-\frac{2m^2}{p^2}\right)
\frac{2p/m}{\sqrt{4-p^2/m^2}}\right.\nonumber\\
&&{\times}\left.\left.\arctan\frac{p/m}{\sqrt{4-p^2/m^2}}\right]\right\},
\end{eqnarray}
where the subscript ``$R$'' represents the radiative correction.
At low-energy limit, $p^2{\rightarrow}0$, this reduces to
\begin{eqnarray}
\omega_R (p,b)|_{p^2{\rightarrow}0}=\frac{2\alpha}{\pi}
\left[\left(\frac{1}{30}+\frac{2}{15}\frac{b^2}{m^2}\right)
\frac{p^2}{m^2}-\frac{2}{15}\frac{\left(p{\cdot}b\right)^2}{m^4}\right].
\end{eqnarray}
The screening of the electric charge produced by the vacuum 
polarization is
\begin{eqnarray}
e^2\longrightarrow \frac{e^2}{1+\omega_R (p,b)}.
\end{eqnarray}

\section{One-loop on-shell vertex}

As in standard QED, the one-loop 
on-shell vertex is
\begin{eqnarray}
-ie\mu^{\epsilon} \bar{u}(q)\Lambda (p,q) u(p)&=&\bar{u}(q)
\int\frac{d^nk}{(2\pi)^n}\left\{(-ie\gamma_\rho\mu^{\epsilon})
\frac{i}{k\hspace{-2.3mm}/+q\hspace{-2mm}/-m-b\hspace{-2mm}/\gamma_5}
(-ie\gamma_\mu\mu^{\epsilon})\right.\nonumber\\
&\times &\left.\frac{i}
{k\hspace{-2.3mm}/+p\hspace{-2mm}/-m-b\hspace{-2mm}/\gamma_5}
(-ie\gamma_\nu\mu^{\epsilon})\frac{-i}{k^2}\left[g_{\nu\rho}-(1-\alpha)
\frac{k_\nu k_\rho}{k^2}\right]
\right\}u(p)\nonumber\\
&{\equiv}&-ie\mu^{\epsilon} \bar{u}(q)\left[\Lambda^{(0)} (p,q)
+\Lambda^{(1)}(p,q) +\Lambda^{(2)}(p,q) \right] u(p),
\label{eq23}
\end{eqnarray}
where we have used the identity Eq.\,(\ref{eq6}) truncated to
 second order in $b$ and $\epsilon$ denotes 
$\epsilon_{\rm UV}$ or $-\epsilon_{\rm IR}$ . 
The relevant Feynman diagrams are shown in
Fig.\,3. Since the radiative correction is gauge
independent, we first have a look at the
gauge dependent part of the amplitude (\ref{eq23}). Eq.\,(\ref{eq23}) 
shows that that
the term proportional to $(1-\alpha)$ involves an integral with integrand
\begin{eqnarray}
\frac{1}{k^4}k\hspace{-2.3mm}/\frac{1}{k\hspace{-2.3mm}/+q\hspace{-2mm}/-m}
({\cdots})\frac{1}{k\hspace{-2.3mm}/+p\hspace{-2mm}/-m}k\hspace{-2.3mm}/.
\end{eqnarray}
Writing the  $k\hspace{-2.3mm}/$ on the left-side as 
$k\hspace{-2.3mm}/+q\hspace{-2mm}/-m-(q\hspace{-2mm}/-m)$
and the  $k\hspace{-2.3mm}/$ on the right-side as 
$k\hspace{-2.3mm}/+p\hspace{-2mm}/-m-(p\hspace{-2mm}/-m)$ and considering
the mass-shell condition, $\bar{u}(q)(q\hspace{-2mm}/-m)=0$,
$(p\hspace{-2mm}/-m)u(p)=0$, one can see that gauge dependent part
is independent of the external momenta and hence is absorbed by the
counterterm of the vertex renormalization, $Z_1^{-1}-1$. 
Therefore, we can calculate the radiative correction of the vertex 
in Feynman gauge, $\alpha=1$\cite{ref16}. 
 
\subsection{One-loop On-shell Vertex in Standard Quantum Electrodynamics} 

$\bar{u}(q)\Lambda^{(0)}_{\mu} (p,q) u(p)$ is the on-shell vertex correction
in standard QED (Fig.\,3a). Using
\begin{eqnarray}
\bar{u}(q)\gamma_\nu (k\hspace{-2mm}/+q\hspace{-2mm}/+m)&=&
\bar{u}(q)\left[-k\hspace{-2mm}/\gamma_\nu+2 (k+q)_\nu\right];\nonumber\\
 (k\hspace{-2mm}/+p\hspace{-2mm}/+m)\gamma_\nu u(p)&=&
\left[-\gamma_\nu k\hspace{-2mm}/+2 (k+p)_\nu\right]u(p),
\end{eqnarray}
we get
\begin{eqnarray}
\bar{u}(q)\Lambda^{(0)}_{\mu} (p,q) u(p)&=&
-ie^2\mu^{2\epsilon}\bar{u}(q)\int\frac{d^nk}{(2\pi)^n}
\frac{\left[-k\hspace{-2mm}/\gamma_\nu+2(k+q)_\nu\right]\gamma_\mu
\left[-\gamma_\nu k\hspace{-2mm}/+2 (k+p)_\nu\right]}
{k^2[(k+q)^2-m^2][(k+p)^2-m^2]}\nonumber\\
&=&-ie^2\mu^{2\epsilon}\bar{u}(q)\int\frac{d^nk}{(2\pi)^n}\left\{
\frac{(2-n)(2k\hspace{-2mm}/k_\mu-k^2\gamma_\mu)}
{k^2[(k+q)^2-m^2][(k+p)^2-m^2]}
\right.\nonumber\\
&&\left.+4\frac{\gamma_\mu k{\cdot}(p+q)+m k_\mu-k\hspace{-2mm}/(p_\mu+q_\mu)
+p{\cdot}q\gamma_\mu}
{k^2[(k+q)^2-m^2][(k+p)^2-m^2]}\right\}u(p).
\end{eqnarray}
Given  that
\begin{eqnarray}
&&\mu^{-2\epsilon_{\rm IR}}\int\frac{d^n k}{(2\pi)^n}
\frac{1}{k^2[(k+q)^2-m^2][(k+p)^2-m^2]}|_{p^2=q^2=m^2,
\epsilon_{\rm IR}\rightarrow 0}
\nonumber\\
&=&-\frac{i}{32\pi^2}\int_0^1dx\frac{1}{m^2-l^2x (1-x)}
\left[\frac{1}{\epsilon_{\rm IR}}+\gamma +
\ln\frac{m^2}{4\pi\mu^2}+\ln\left(1-\frac{l^2}{m^2}x (1-x)
\right)\right]\nonumber\\
&=&-\frac{i}{32\pi^2}\frac{1}{m^2}\left[\left(\frac{1}{\epsilon_{\rm IR}}+
\ln\frac{m^2}{4\pi\mu^2}+\gamma\right)\frac{4m}{l\sqrt{4-l^2/m^2}}
\arctan\frac{l/m}{\sqrt{4-l^2/m^2}}\right.\nonumber\\
&& \left.+\int_0^1 dx \frac{\ln (1-l^2/m^2 x (1-x)}{1-l^2/m^2 x (1-x)}\right];
\end{eqnarray}
\begin{eqnarray}
&&\int \frac{d^nk}{(2\pi)^n}\frac{k_{\mu}}{k^2[(k+q)^2-m^2][(k+p)^2-m^2]}
|_{p^2=q^2=m^2}\nonumber\\
&=&\frac{i}{8\pi^2}\frac{p_\mu+q_\mu}{m^2}\frac{m}{l\sqrt{4-l^2/m^2}}
\arctan\frac{l/m}{\sqrt{4-l^2/m^2}};
\end{eqnarray}
\begin{eqnarray}
&&\mu^{2\epsilon_{\rm UV}}\int \frac{d^nk}{(2\pi)^n}\frac{k_{\mu}k_{\nu}}
{k^2[(k+q)^2-m^2][(k+p)^2-m^2]}
|_{p^2=q^2=m^2,\epsilon_{\rm UV}\rightarrow 0}\nonumber\\
&=&\frac{i}{64\pi^2}g_{\mu\nu}\int_0^1dx
\left[\frac{1}{\epsilon_{\rm UV}}-\gamma
+\ln\frac{4\pi\mu^2}{m^2}-\ln\left(1-\frac{l^2}{m^2} x (1-x)\right)
\right]\nonumber\\
&&-\frac{i}{32\pi^2}\int_0^1dx
\frac{1}{m^2-l^2x (1-x)}\left[x (1-x)\left(p_\mu q_\nu+p_\nu q_\mu\right)
+x^2 \left(p_\mu p_\nu+q_\mu q_\nu\right)\right]\nonumber\\
&=&\frac{i}{64\pi^2}g_{\mu\nu}\left(\frac{1}{\epsilon_{\rm UV}}-\gamma+3
+\ln\frac{4\pi\mu^2}{m^2}-\frac{2m}{l}\sqrt{4-l^2/m^2}
\arctan\frac{l/m}{\sqrt{4-l^2/m^2}}\right]\nonumber\\
&-&\frac{i}{32\pi^2}\frac{p_\mu q_\nu+p_\nu q_\mu}{m^2}
\left[-\frac{m^2}{l^2}+\frac{4m^3}{l^3}\frac{1}{\sqrt{4-l^2/m^2}}
\arctan\frac{l/m}{\sqrt{4-l^2/m^2}}\right]\nonumber\\
&-&\frac{i}{32\pi^2}\frac{p_\mu p_\nu+q_\mu q_\nu}{m^2}
\left[\frac{m^2}{l^2}+2\frac{m}{l}\left(1-2\frac{m^2}{l^2}\right)
\frac{1}{\sqrt{4-l^2/m^2}}\arctan\frac{l/m}{\sqrt{4-l^2/m^2}}\right],
\end{eqnarray}
we obtain the result in dimensional regularization
\begin{eqnarray}
&&\bar{u}(q)\Lambda^{(0)}_{\mu} (p,q) u(p)\nonumber\\
&=&\bar{u}(q)\left\{\frac{e^2}{16\pi^2}
\gamma_\mu\left(\frac{1}{\epsilon_{\rm UV}}-\gamma
+\ln\frac{4\pi\mu^2}{m^2}+\frac{6m}{l}\sqrt{4-l^2/m^2}
\arctan\frac{l/m}{\sqrt{4-l^2/m^2}}\right)\right.\nonumber\\
&+&\frac{e^2}{2\pi^2}\frac{p_\mu+q_\mu}{m}\frac{m}{l}\left(1-
\frac{1}{\sqrt{4-l^2/m^2}}\right)\arctan\frac{l/m}{\sqrt{4-l^2/m^2}}\nonumber\\
&-&\frac{e^2}{16\pi^2}\gamma_\mu\left(2-\frac{l^2}{m^2}\right)
\left[\left(\frac{1}{\epsilon_{\rm IR}}-\ln\frac{4\pi\mu^2}{m^2}+\gamma\right)
\frac{4m/l}{\sqrt{4-l^2/m^2}}\arctan\frac{l/m}{\sqrt{4-l^2/m^2}}\right.
\nonumber\\
&+&\left.\left.\int_0^1 dx
\frac{\ln (1-l^2/m^2 x (1-x)}{1-l^2/m^2 x (1-x)}\right]\right\}u(p).
\label{eq30a}
\end{eqnarray}
In above equations, $l{\equiv}q-p$,
to distinguish the infrared and ultraviolet divergences, we especially
put on the subscripts on the regulators to emphasize
the difference, $\epsilon_{\rm IR}=n/2-2$ and
$\epsilon_{\rm UV}=2-n/2$.

\subsection{On-shell Vertex Correction at First Order of $b$}

Now let us look at the contribution from the first order of $b_\mu$ (Fig.\,3b),
\begin{eqnarray}
\Lambda^{(1)}_{\mu} (p,q) 
&=&-ie^2\mu^{-2\epsilon_{\rm IR}}\int\frac{d^nk}{(2\pi)^n}
\frac{1}{k^2}\left[\gamma_\nu\frac{1}{k\hspace{-2.3mm}/+q\hspace{-2mm}/-m}
 \gamma_\mu \frac{1}{k\hspace{-2.3mm}/+p\hspace{-2mm}/-m}b\hspace{-2mm}/
\gamma_5\frac{1}{k\hspace{-2.3mm}/+p\hspace{-2mm}/-m}
\gamma_\nu\right.\nonumber\\
&&+\left.\gamma_\nu\frac{1}{k\hspace{-2.3mm}/+q\hspace{-2mm}/-m}b\hspace{-2mm}/
\gamma_5\frac{1}{k\hspace{-2.3mm}/+q\hspace{-2mm}/-m}\gamma_\mu
\frac{1}{k\hspace{-2.3mm}/+p\hspace{-2mm}/-m}\gamma_\nu\right]\nonumber\\
&=& -ie^2\mu^{-2\epsilon_{\rm IR}}\int\frac{d^nk}{(2\pi)^n}\frac{1}{k^2}
\left\{-b^{\alpha}\frac{\partial}{\partial p^\alpha}\left[
\gamma_\nu\frac{1}{k\hspace{-2.3mm}/+q\hspace{-2mm}/-m}
 \gamma_\mu \frac{1}{k\hspace{-2.3mm}/+p\hspace{-2mm}/-m}\gamma_\nu\gamma_5
\right]\right.\nonumber\\
&&+b^{\alpha}\frac{\partial}{\partial p^\alpha}\left[\gamma_5\gamma_\nu
\frac{1}{k\hspace{-2.3mm}/+q\hspace{-2mm}/-m}\gamma_\mu
\frac{1}{k\hspace{-2.3mm}/+p\hspace{-2mm}/-m}\gamma_\nu\right]\nonumber\\
&&-\frac{2m}{k^2[(k+p)^2-m^2]}\gamma_\nu
\frac{1}{k\hspace{-2.3mm}/+q\hspace{-2mm}/-m}
\gamma_\mu \frac{1}{k\hspace{-2.3mm}/+p\hspace{-2mm}/-m}
b\hspace{-2mm}/\gamma_\nu\gamma_5\nonumber\\
&&\left.+\frac{2m}{k^2[(k+q)^2-m^2]}
\gamma_5\gamma_\nu b\hspace{-2mm}/
\frac{1}{k\hspace{-2.3mm}/+q\hspace{-2mm}/-m}\gamma_\mu
\frac{1}{k\hspace{-2.3mm}/+p\hspace{-2mm}/-m}\gamma_\nu\right\}.
\end{eqnarray}
After performing some algebraic operation and taking the derivative, we have
\begin{eqnarray}
\Lambda^{(1)}_{\mu} (p,q) 
&=&-ie^2\mu^{-2\epsilon_{\rm IR}}\int\frac{d^nk}{(2\pi)^n}\left\{
\frac{-\gamma_\nu (k\hspace{-2.3mm}/+q\hspace{-2mm}/+m)
\gamma_\mu b\hspace{-2mm}/\gamma_\nu\gamma_5
+\gamma_5\gamma_\nu b\hspace{-2mm}/\gamma_\mu 
(k\hspace{-2.3mm}/+p\hspace{-2mm}/+m)\gamma_\nu}{k^2[(k+p)^2-m^2][(k+p)^2-m^2]}
\right. 
\nonumber\\
&&+\frac{2b{\cdot}(k+p)\left[\gamma_\nu (k\hspace{-2.3mm}/+q\hspace{-2mm}/+m)
\gamma_\mu (k\hspace{-2.3mm}/+p\hspace{-2mm}/+m)\gamma_\nu\gamma_5\right]}
{k^2[(k+p)^2-m^2]^2[(k+q)^2-m^2]} 
\nonumber\\
&&-\frac{2b{\cdot}(k+q)
\left[\gamma_5\gamma_\nu (k\hspace{-2.3mm}/+q\hspace{-2mm}/+m)
\gamma_\mu (k\hspace{-2.3mm}/+p\hspace{-2mm}/+m)\gamma_\nu\right]}
{k^2[(k+p)^2-m^2]^2[(k+q)^2-m^2]^2} 
\nonumber\\
&&-\frac{2m\left[\gamma_\nu (k\hspace{-2.3mm}/+q\hspace{-2mm}/+m)
\gamma_\mu (k\hspace{-2.3mm}/+p\hspace{-2mm}/+m)\gamma_\nu\gamma_5\right]}
{k^2[(k+p)^2-m^2]^2[(k+q)^2-m^2]} 
\nonumber\\
&&+\left.\frac{2m\left[\gamma_5\gamma_\nu (k\hspace{-2.3mm}/+q\hspace{-2mm}/+m)
\gamma_\mu (k\hspace{-2.3mm}/+p\hspace{-2mm}/+m)\gamma_\nu\right]}
{k^2[(k+p)^2-m^2]^2[(k+q)^2-m^2]^2}\right\}.
\label{eq32}
\end{eqnarray}
Putting Eq.\,(\ref{eq32}) on mass-shell, i.e. evaluating 
$\bar{u}(q)\Lambda^{(1)}_{\mu} (p,q) u(p)$, and making use of
the following $\gamma$ matrices formula,
\begin{eqnarray}
&&\bar{u}(q)[\gamma_\nu (k\hspace{-2.3mm}/+q\hspace{-2mm}/+m)
\gamma_\mu (k\hspace{-2.3mm}/+p\hspace{-2mm}/+m)\gamma_\nu\gamma_5] u(p)
=\left[4k\cdot (p+q)-2 l^2+2m k\hspace{-2.3mm}/
+2 k^2\right]\gamma_\mu\gamma_5 \nonumber\\
&&+\left[-4 (p_\mu+q_\mu) k\hspace{-2.3mm}/
+6m\gamma_\mu k\hspace{-2.3mm}/
+8m q_\mu-4  k\hspace{-2.3mm}/k_\mu\right]\gamma_5
+\epsilon_{\rm IR}\left(4m k\hspace{-2.3mm}/\gamma_\mu
-4k\hspace{-2.3mm}/k_\mu+2 k^2\gamma_\mu\right)\gamma_5;
\nonumber\\
&&\bar{u}(q)\left[\gamma_5\gamma_\nu (k\hspace{-2.3mm}/+q\hspace{-2mm}/+m)
\gamma_\mu (k\hspace{-2.3mm}/+p\hspace{-2mm}/+m)\gamma_\nu\right]u(p)
=\gamma_5\gamma_\mu\left[4k\cdot (p+q)-2 l^2+2m k\hspace{-2.3mm}/+2 k^2\right]
\nonumber\\
&&+\gamma_5\left[-4 (p_\mu+q_\mu)
k\hspace{-2.3mm}/+6mk\hspace{-2.3mm}/\gamma_\mu
+8m p_\mu-4k\hspace{-2.3mm}/k_\mu\right]
+\epsilon_{\rm IR}\gamma_5\left(4m\gamma_\mu k\hspace{-2.3mm}/-4 
 k\hspace{-2.3mm}/k_\mu+2\gamma_\mu k^2 \right);\nonumber\\
&&\bar{u}(q)[\gamma_\nu (k\hspace{-2.3mm}/+q\hspace{-2mm}/+m)
\gamma_\mu (k\hspace{-2.3mm}/+p\hspace{-2mm}/+m)b\hspace{-2mm}/\gamma_\nu
\gamma_5] u(p)=\left(8k_\mu+8q_\mu-4m\gamma_\mu\right)
(k+p)\cdot b\gamma_5\nonumber\\
&&-[4k\cdot (p+q)+4p\cdot q+2 mk\hspace{-2.3mm}/]\gamma_\mu b\hspace{-2mm}/
\gamma_5+(2m\gamma_\mu-4q_\mu)b\hspace{-2mm}/k\hspace{-2.3mm}/\gamma_5
+(4q\cdot b\gamma_\mu k\hspace{-2.3mm}/-4p\cdot b k\hspace{-2.3mm}/\gamma_\mu
\nonumber\\
&&+4p_\mu k\hspace{-2.3mm}/ b\hspace{-2mm}/-4k^2b_\mu)\gamma_5
-\epsilon_{\rm IR}[4(k+p)\cdot b k\hspace{-2.3mm}/\gamma_\mu+2
 k\hspace{-2.3mm}/\gamma_\mu  b\hspace{-2mm}/ k\hspace{-2.3mm}/]\gamma_5;
\nonumber\\
&&\bar{u}(q)[\gamma_5\gamma_\nu  b\hspace{-2mm}/(k\hspace{-2.3mm}/
+q\hspace{-2mm}/+m)
\gamma_\mu (k\hspace{-2.3mm}/+p\hspace{-2mm}/+m)\gamma_\nu] u(p)
=\gamma_5(k+q)\cdot b\left(8k_\mu + 8p_\mu-4m\gamma_\mu\right)\nonumber\\
&&-\gamma_5 b\hspace{-2mm}/\gamma_\mu[4k\cdot (p+q)+4 p\cdot q-2m
k\hspace{-2mm}/]+\gamma_5k\hspace{-2.3mm}/b\hspace{-2mm}/
(2 m\gamma_\mu-4p_\mu)+\gamma_5(4p\cdot b k\hspace{-2.3mm}/\gamma_\mu
+4q_\mu b\hspace{-2mm}/k\hspace{-2mm}/\nonumber\\
&&-4q\cdot b\gamma_\mu k\hspace{-2mm}/-4b_\mu k^2)k\hspace{-2mm}/
+\epsilon_{\rm IR}\gamma_5[4b\cdot (k+q)\gamma_\mu k\hspace{-2mm}/
-2k\hspace{-2mm}/b\hspace{-2mm}/\gamma_\mu k\hspace{-2mm}/],
\label{eq32a1}
\end{eqnarray}
we obtain
\begin{eqnarray}
&&\bar{u}(q)\Lambda^{(1)} (p,q) u(p)=\bar{u}(q)\left\{
\frac{e^2}{4\pi^2}\int_0^1dx \frac{1}{m^2-l^2 x(1-x)}\left[
\left(1-\frac{9}{2}x\right)(p+q)\cdot b\gamma_\mu \right.\right.\nonumber\\
&&\left.+\frac{5}{2}x (p_\mu+q_\mu)b\hspace{-2mm}/
+m\left(\frac{1}{8}-x\right)[b\hspace{-2mm}/,\gamma_\mu]
-\frac{1}{4}mb_\mu\right]\gamma_5
\nonumber\\
&&-\frac{e^2}{8\pi^2}\int_0^1dx \frac{1}{m^2-l^2 x(1-x)}\left[
\frac{1}{\epsilon_{\rm IR}}+\ln\frac{m^2}{4\pi\mu^2}
+\ln\left(1-\frac{l^2}{m^2}x (1-x)\right)+\gamma\right]\nonumber\\
&&\times \left[-(p+q)\cdot b\gamma_\mu 
+(p_\mu+q_\mu)b\hspace{-2mm}/+\frac{1}{2}m[b\hspace{-2mm}/,\gamma_\mu]\right]
\gamma_5\nonumber\\
&&+\frac{e^2}{4\pi^2}\int_0^1dx \frac{x}{[m^2-l^2 x(1-x)]^2}\left[
(p+q)\cdot b\gamma_\mu (m^2-l^2)+
m^2 (p_\mu+q_\mu)b\hspace{-2mm}/\right.\nonumber\\
&&\left.-\frac{ml^2}{2}[b\hspace{-2mm}/,\gamma_\mu]
+m (p_\mu+q_\mu)l\cdot b(3+6 x-5 x^2)+l_\mu m(p+q)\cdot b(-1+3 x+5 x^2)
\right]\gamma_5
\nonumber\\
&&+\frac{e^2}{4\pi^2}\int_0^1dx \frac{x}{[m^2-l^2 x(1-x)]^2}\left[
\frac{1}{\epsilon_{\rm IR}}+\ln\frac{m^2}{4\pi\mu^2}
+\ln\left(1-\frac{l^2}{m^2}x (1-x)\right)+\gamma\right]\nonumber\\
&&\times \left[2(l^2-2m^2)(p+q)\cdot b\gamma_\mu 
+m^2(p_\mu+q_\mu)b\hspace{-2mm}/+\frac{1}{2}m(4m^2-l^2)
[b\hspace{-2mm}/,\gamma_\mu]\right.\nonumber\\
&&\left.\left.+\frac{1}{2}(p_\mu+q_\mu)m l\cdot b(16 x-11)
+\frac{1}{2}l_\mu m (p+q)\cdot b (1+2 x)\right]\gamma_5\right\}u(p).
\label{eq32a2}
\end{eqnarray}
With the Gordon identity (\ref{eqgor2}), the superficial
antisymmetric terms of Eq.\,(\ref{eq32a2}) in $p$, $q$ such 
as $(p_\mu+q_\mu)l\cdot b\gamma_5$
and $l_\mu (p+q)\cdot b$ can be converted into
an explicit $p$, $q$ symmetric form. For examples, there have
\begin{eqnarray}
\bar{u}(q)(p_\mu+q_\mu)l\cdot b \gamma_5 u(p)&=&
\bar{u}(q)(p+q)_\mu\left[2m b\hspace{-2mm}/\gamma_5
+\frac{1}{2}\epsilon_{\nu\rho\alpha\beta}b^{\nu}(p+q)^{\rho}
\sigma^{\alpha\beta}\right]u(p),\nonumber\\
\bar{u}(q)(p+q)\cdot bl_\mu \gamma_5 u(p)&=&
\bar{u}(q)(p+q)\cdot b\left[2m\gamma_\mu\gamma_5
+\frac{1}{2}\epsilon_{\mu\nu\alpha\beta}(p+q)^{\nu}\sigma^{\alpha\beta}\right]
u(p).
\end{eqnarray}

\subsection{ Contribution to Second Order in $b_\mu$}

The on-shell quantum vertex at second order in $b_\mu$ is quite 
complicated. It gets  contributions from three possible insertions
of $b\hspace{-2mm}/\gamma_5$ vertex as shown in Fig.\,3c,
\begin{eqnarray}
\bar{u}(q)\Lambda^{(2)}_{\mu} (p,q) u(p)&=&
-ie^2\mu^{-2\epsilon_{\rm IR}}\bar{u}(q)\int\frac{d^nk}{(2\pi)^n}
\frac{1}{k^2}\left[\gamma_\nu\frac{1}{k\hspace{-2.3mm}/+q\hspace{-2mm}/-m}
\gamma_\mu \frac{1}{k\hspace{-2.3mm}/+p\hspace{-2mm}/-m}b\hspace{-2mm}/
\gamma_5 \right.\nonumber\\
&&{\times}\frac{1}{k\hspace{-2.3mm}/+p\hspace{-2mm}/-m}b\hspace{-2mm}/
\gamma_5 \frac{1}{k\hspace{-2.3mm}/+p\hspace{-2mm}/-m}\gamma_\nu
\nonumber\\
&&+\gamma_\nu\frac{1}{k\hspace{-2.3mm}/+q\hspace{-2mm}/-m}
b\hspace{-2mm}/\gamma_5\frac{1}{k\hspace{-2.3mm}/+q\hspace{-2mm}/-m}
b\hspace{-2mm}/\gamma_5\frac{1}{k\hspace{-2.3mm}/+q\hspace{-2mm}/-m}
\gamma_\mu \frac{1}{k\hspace{-2.3mm}/+p\hspace{-2mm}/-m}\gamma_\nu\nonumber\\
&&+\left.\gamma_\nu\frac{1}{k\hspace{-2mm}/+q\hspace{-2mm}/-m}
b\hspace{-2mm}/\gamma_5\frac{1}{k\hspace{-2.3mm}/+q\hspace{-2mm}/-m}
\gamma_\mu \frac{1}{k\hspace{-2.3mm}/+p\hspace{-2mm}/-m}
b\hspace{-2mm}/\gamma_5\frac{1}{k\hspace{-2.3mm}/+p\hspace{-2mm}/-m}
\gamma_\nu\right]u(p).
\label{eq34}
\end{eqnarray}
To reduce the number of complicated  $\gamma$-matrix  operations, we continue
to use  the identities (\ref{eq13}) and (\ref{eq14}). As a consequence,
Eq.\,(\ref{eq34}) can be rewritten in the following compact form,
\begin{eqnarray}
\bar{u}(q)\Lambda^{(2)}_{\mu} (p,q) u(p)&=&
-ie^2\mu^{2\epsilon_{\rm IR}}\bar{u}(q)\int\frac{d^nk}{(2\pi)^n}
\frac{1}{k^2}\left\{\frac{1}{2}
\left(b^\alpha\frac{\partial}{\partial p^\alpha}
+b^\alpha\frac{\partial}{\partial q^\alpha}\right)^2\right.\nonumber\\
&&\times\left(\gamma_\nu\frac{1}{k\hspace{-2.3mm}/+q\hspace{-2mm}/-m}
\gamma_\mu\frac{1}{k\hspace{-2.3mm}/+p\hspace{-2mm}/-m}\gamma_\nu\right)
~~~~\bigcirc\hspace{-4mm}1
\nonumber\\
&&-\frac{2mb^2}{(k+p)^2-m^2}
\gamma_\nu\frac{1}{k\hspace{-2.3mm}/+q\hspace{-2mm}/-m}
\gamma_\mu\left(\frac{1}{k\hspace{-2.3mm}/+p\hspace{-2mm}/-m}\right)^2
\gamma_\nu
~~~~\bigcirc\hspace{-4mm}2 \nonumber\\
&&-\frac{2mb^2}{(k+q)^2-m^2}
\gamma_\nu\left(\frac{1}{k\hspace{-2.3mm}/+q\hspace{-2mm}/-m}\right)^2
\gamma_\mu\frac{1}{k\hspace{-2.3mm}/+p\hspace{-2mm}/-m}\gamma_\nu
~~~~\bigcirc\hspace{-4mm}3 \nonumber\\
&&-\frac{2mb^2}{(k+p)^2-m^2}b^\alpha\frac{\partial}{\partial q^\alpha}
\left(\gamma_\nu \frac{1}{k\hspace{-2.3mm}/+q\hspace{-2mm}/-m}
\gamma_\mu b\hspace{-2mm}/\frac{1}{k\hspace{-2.3mm}/+p\hspace{-2mm}/-m}
\gamma_\nu\right)~~~~\bigcirc\hspace{-4mm}4 \nonumber\\
&&-\frac{2mb^2}{(k+q)^2-m^2}b^\alpha\frac{\partial}{\partial p^\alpha}
\left(\gamma_\nu \frac{1}{k\hspace{-2.3mm}/+q\hspace{-2mm}/-m}
b\hspace{-2mm}/\gamma_\mu \frac{1}{k\hspace{-2.3mm}/+p\hspace{-2mm}/-m}
\gamma_\nu\right)~~~~\bigcirc\hspace{-4mm}5 \nonumber\\
&&\hspace{-15mm}\left.+\frac{4m^2}{[(k+p)^2-m^2][(k+q)^2-m^2]}
\gamma_\nu \frac{1}{k\hspace{-2.3mm}/+q\hspace{-2mm}/-m}
b\hspace{-2mm}/\gamma_\mu b\hspace{-2mm}/
\frac{1}{k\hspace{-2.3mm}/+p\hspace{-2mm}/-m}
\gamma_\nu\right\}u(p).~\bigcirc\hspace{-4mm}6
\label{eq35}
\end{eqnarray}
The last term is quite simple,
\begin{eqnarray}
\bigcirc\hspace{-3.5mm}6&=&\bar{u}(q)\left\{\frac{e^2}{4\pi^2}
\int_0^1d x \frac{x (1-x)}{[m^2-l^2 x (1-x)]^2}
\left[\frac{1}{\epsilon_{\rm IR}}+\ln\frac{m^2}{4\pi\mu^2}
+\ln\left(1-\frac{l^2}{m^2}x (1-x)\right)\right.\right.\nonumber\\
&&\left.-1+\gamma\right]m^2\left(
2b\hspace{-2mm}/b_\mu-b^2\gamma_\mu\right)\nonumber\\
&-&\frac{e^2}{2\pi^2}\int_0^1d x \frac{x (1-x)}{[m^2-l^2 x (1-x)]^3}
\left[\frac{2}{\epsilon_{\rm IR}}+
\ln\frac{m^2}{4\pi\mu^2}
+\ln\left(1-\frac{l^2}{m^2}x (1-x)\right)\right.\nonumber\\
&&\left.+5+2\gamma\right] m^2
\left[-2 mb_\mu b{\cdot}(p+q)+ b^2 (p_\mu+q_\mu)\right]\nonumber\\
&+&\left.\frac{e^2}{\pi^2}\int_0^1d x \frac{x (1-x)}{[m^2-l^2 x (1-x)]^3}
m^2\left(-6m^2+l^2\right) \left(2b\hspace{-2mm}/b_\mu-b^2\gamma_\mu\right)
\right\}u(p).
\label{eq36}
\end{eqnarray}
The second and the third terms yield
\begin{eqnarray}
\bigcirc\hspace{-3.5mm}2+\bigcirc\hspace{-3.4mm}3
&=&\bar{u}(q)\left\{\frac{e^2}{4\pi^2}
\int_0^1 d x\frac{x^2}{[m^2-l^2 x(1-x)]^2}3m^2 b^2
\gamma_\mu\right.\nonumber\\
&+&\frac{e^2}{8\pi^2}\int_0^1d x\frac{x^2mb^2 (p_\mu+q_\mu)}
{[m^2-l^2 x(1-x)]^2}
\left[\frac{1}{\epsilon_{\rm IR}}-\ln\frac{m^2}{4\pi\mu^2}
-\ln\left(1-\frac{l^2}{m^2}x (1-x)\right)-3+\gamma\right]
\nonumber\\
&&\hspace{-15mm}-\frac{e^2}{4\pi^2}
\int_0^1d x\frac{x^2(1-x)mb^2l^2(p_\mu+q_\mu)}
{[m^2-l^2 x(1-x)]^3}
\left[\frac{2}{\epsilon_{\rm IR}}+\ln\frac{m^2}{4\pi\mu^2}
+\ln\left(1-\frac{l^2}{m^2}x (1-x)\right)-3+2\gamma\right]
\nonumber\\
&+&\frac{e^2}{2\pi^2}\int_0^1d x\frac{x^2 mb^2}{[m^2-l^2 x(1-x)]^3}
\left[ (p_\mu+q_\mu)\left(-2 m^2+l^2 (2-x)\right)(1-x)\right.\nonumber\\
&&\left.\left.+m\gamma_\mu\left(-16m^2+4 m^2x
+4 l^2-l^2x+2 l^2 x^2\right)\right]\right\}u(p). 
\label{eq37}
\end{eqnarray}
As for the fourth and fifth terms,
\begin{eqnarray}
\bigcirc\hspace{-3.5mm}4+\bigcirc\hspace{-3.4mm}5
&=&-ie^2\mu^{-2\epsilon_{\rm IR}}\bar{u}(q)\left\{-\int \frac{d^nk}{(2\pi)^n}
\frac{2m\gamma_\nu b\hspace{-2mm}/
\gamma_\mu b\hspace{-2mm}/(k\hspace{-2.3mm}/+p\hspace{-2mm}/+m)\gamma_\nu}
{k^2[(k+p)^2-m^2]^2[(k+q)^2-m^2]}\right.\nonumber\\
&&-\frac{2m\gamma_\nu (k\hspace{-2.3mm}/+q\hspace{-2mm}/+m)b\hspace{-2mm}/
\gamma_\mu b\hspace{-2mm}/\gamma_\nu}{k^2[(k+p)^2-m^2][(k+q)^2-m^2]^2}
\nonumber\\
&&+\frac{4mb{\cdot}(k+q)\gamma_\nu(k\hspace{-2.3mm}/+q\hspace{-2mm}/+m)
\gamma_\mu b\hspace{-2mm}/(k\hspace{-2.3mm}/+p\hspace{-2mm}/+m)\gamma_\nu}
{k^2[(k+p)^2-m^2]^2[(k+q)^2-m^2]^2}\nonumber\\
&&+\left.\frac{4mb{\cdot}(k+p)\gamma_\nu(k\hspace{-2.3mm}/+q\hspace{-2mm}/+m)
 b\hspace{-2mm}/\gamma_\mu (k\hspace{-2.3mm}/+p\hspace{-2mm}/+m)\gamma_\nu}
{k^2[(k+p)^2-m^2]^2[(k+q)^2-m^2]^2}\right\}u(p),
\label{eq38}
\end{eqnarray}
a straightforward  calculation leads to
\begin{eqnarray}
\bigcirc\hspace{-3.5mm}4+\bigcirc\hspace{-3.4mm}5
&=&\bar{u}(q)\left\{\frac{e^2}{4\pi^2}\int_0^1d x \frac{x}{[m^2-l^2 x(1-x)]^2}
m\left[4b_\mu b{\cdot}(p+q)-b^2 (p_\mu+q_\mu)\right]\right.\nonumber\\
&&\times
\left[\frac{1}{\epsilon_{\rm IR}}+\ln\frac{m^2}{4\pi^2}+
\ln\left(1-\frac{l^2}{m^2}x (1-x)\right)+2-x+\gamma\right]\nonumber\\
&+&\frac{e^2}{4\pi^2}\int_0^1d x \frac{mx}{[m^2-l^2 x(1-x)]^2}
\left[m\left(4b_\mu b\hspace{-2mm}/-b^2\gamma_\mu\right)(2x-3)
-4(1-x) b{\cdot}(p+q)b_\mu\right] \nonumber\\
&+&\frac{e^2}{2\pi^2}\int_0^1d x \frac{x(1-x)}{[m^2-l^2 x(1-x)]^2}
\left[\frac{1}{\epsilon_{\rm IR}}+\ln\frac{m^2}{4\pi^2}+
\ln\left(1-\frac{l^2}{m^2}x (1-x)\right)\right.\nonumber\\
&&\left.-1+\gamma\right]
\left[3mb{\cdot}(p+q)b_\mu-m^2b\hspace{-2mm}/b_\mu\right]
\nonumber\\
&-&\frac{e^2}{2\pi^2}\int_0^1d x \frac{x(1-x)}{[m^2-l^2 x(1-x)]^3}
\left[\frac{2}{\epsilon_{\rm IR}}+\ln\frac{m^2}{4\pi^2}\right.\nonumber\\
&&\left.+\ln\left(1-\frac{l^2}{m^2}x (1-x)\right)
-3+2\gamma\right]m\left(2m^2-l^2\right) b{\cdot}(p+q) b_\mu\nonumber\\
&-&\frac{2e^2}{\pi^2}\int_0^1d x \frac{x(1-x)}{[m^2-l^2 x(1-x)]^3}
\left[m(2m^2-l^2)(p+q){\cdot}bb_\mu\right.\nonumber\\
&&\left.\left.-m^2\left[(p\cdot b)^2+
(p\cdot b)^2\right]+m^2(p{\cdot}bp_\mu+q{\cdot}bq_\mu)b\hspace{-2mm}/
-m^2(p{\cdot}bq_\mu+p{\cdot}bp_\mu)b\hspace{-2mm}/\right]\right\}u(p).
\label{eq39}
\end{eqnarray}
The first term is the most complicated. When the derivative is 
taken, it becomes 
 \begin{eqnarray}
\bigcirc\hspace{-3.4mm}1
&=&-ie^2\mu^{-2\epsilon_{\rm IR}}\bar{u}(q)
\int\frac{d^nk}{(2\pi)^n}\left\{\frac{\gamma_\nu b\hspace{-2mm}/\gamma_\mu
b\hspace{-2mm}/\gamma_\nu}{k^2[(k+p)^2-m^2][(k+q)^2-m^2]}\right.\nonumber\\
&-&\frac{1}{k^2}\left[\frac{2b{\cdot}(k+p)}{[(k+p)^2-m^2]^2[(k+q)^2-m^2]}
+\frac{2b{\cdot}(k+q)}{[(k+p)^2-m^2][(k+q)^2-m^2]^2}\right]\nonumber\\
&&{\times}\left[\gamma_\nu (k\hspace{-2mm}/+q\hspace{-2mm}/+m)\gamma_\mu
b\hspace{-2mm}/\gamma_\nu+\gamma_\nu b\hspace{-2mm}/\gamma_\mu
(k\hspace{-2mm}/+p\hspace{-2mm}/+m)\gamma_\nu\right]\nonumber\\
&-&\frac{1}{k^2}\left[\frac{b^2}{[(k+p)^2-m^2]^2[(k+q)^2-m^2]}
+\frac{b^2}{[(k+p)^2-m^2][(k+q)^2-m^2]^2}\right]\nonumber\\
&&\times \left[
\gamma_\nu (k\hspace{-2mm}/+q\hspace{-2mm}/+m)\gamma_\mu
(k\hspace{-2mm}/+p\hspace{-2mm}/+m)\gamma_\nu\right]\nonumber\\
&+&\frac{4}{k^2}\left[\frac{[b{\cdot}(k+p)]^2}
{[(k+p)^2-m^2]^3[(k+q)^2-m^2]}
+\frac{[b{\cdot}(k+q)]^2}{[(k+p)^2-m^2][(k+q)^2-m^2]^3}\right.\nonumber\\
&&\left.\left.+\frac{[b{\cdot}(k+p)][b{\cdot}(k+q)]}{[(k+p)^2-m^2]^2
[(k+q)^2-m^2]^2}\right]
\left[\gamma_\nu (k\hspace{-2mm}/+q\hspace{-2mm}/+m)\gamma_\mu
(k\hspace{-2mm}/+p\hspace{-2mm}/+m)\gamma_\nu\right]\right\}u(p).
\label{eq40}
\end{eqnarray} 
After Feynman parameterization and the momentum integration, we get
 \begin{eqnarray}
\bigcirc\hspace{-3.4mm}1
&=&\bar{u}(q)\left(\frac{e^2}{8\pi^2}b^2
\gamma_\mu\int_0^1 dx\frac{x^2-2 x-1}{m^2-l^2 x(1-x)}
-\frac{i}{4\pi^2}b_\mu b\hspace{-2mm}/\int_0^1 dx\frac{x^2+2 x-1}
{m^2-l^2 x(1-x)}\right.  \nonumber\\
&+&\frac{e^2}{4\pi^2}
\int_0^1 dx \frac{mb_\mu (p+q){\cdot}b}{[m^2-l^2 x(1-x)]^2}
\left\{x^2-3 x+2 x\left[\frac{1}{\epsilon_{\rm IR}}+\ln\frac{m^2}{4\pi\mu^2}
\right.\right.\nonumber\\
&&\left.\left.+\ln\left(1-\frac{l^2}{m^2}x (1-x)\right)+\gamma\right]
\right\}\nonumber\\
&+&\frac{e^2}{4\pi^2}\gamma_\mu \int_0^1 dx \frac{1}{[m^2-l^2 x(1-x)]^2}
\left\{-\frac{1}{2}[b\cdot (p+q)]^2 (x^2+6 x)\right.\nonumber\\
&&-\frac{1}{2}(l\cdot b)^2(x-4 x^2-2 x^3+4 x^4)- m^2 b^2 (1+x^2)
+\frac{1}{2}l^2 b^2 (x^2+x+1)\nonumber\\
&&+\left(\frac{1}{\epsilon_{\rm IR}}+\gamma\right)
\left[\frac{7}{4}x(p{\cdot}b+q{\cdot}b)^2
+\frac{1}{4}(x-2 x^2) (l\cdot b)^2
+m^2 b^2 (1-x)^2\right.\nonumber\\
&&\left. +\frac{1}{2}l^2 b^2 (-1+x-x^2)\right]
+\left[\ln\frac{m^2}{4\pi\mu^2}
+\ln\left(1-\frac{l^2}{m^2}x (1-x)\right)\right]
\left[\frac{9}{4}x (p\cdot b+q\cdot b)^2 
\right.\nonumber\\
&&\left.\left.+m^2 b^2 (1-x)^2+\frac{1}{2}l^2 b^2 
(-1+x-x^2)+\frac{1}{4}(l\cdot b)^2 (2 x^2-x)\right]\right\}\nonumber\\
&+&\frac{e^2}{4\pi^2}\int_0^1 dx \frac{b\hspace{-2mm}/}{[m^2-l^2 x(1-x)]^2}
\left\{b{\cdot}(p+q)(p_\mu+q_\mu)\frac{1}{2}(x^2+4x)\right.\nonumber\\
&&+\frac{1}{2}l\cdot bl_\mu(-2 x+5 x^2-4 x^3+4 x^4)
+\left[\frac{1}{\epsilon_{\rm IR}}+\ln\frac{m^2}{4\pi\mu^2}
+\ln\left(1-\frac{l^2}{m^2}x (1-x)\right)+\gamma\right]\nonumber\\
&&\times \left.\left[-\frac{5}{2} (p+q){\cdot}b (p_\mu+q_\mu)x+\frac{1}{2}
(x-2x^2)l\cdot bl_\mu\right]\right\}\nonumber\\
&+&\frac{e^2}{8\pi^2}\int_0^1 dx \frac{mb^2 (p_\mu+q_\mu)}{[m^2-l^2 x(1-x)]^2}
(x^2-2 x)\nonumber\\
&+&\frac{e^2}{2\pi^2}\int_0^1 dx \frac{m}{[m^2-l^2 x(1-x)]^3}
\left\{l_\mu l{\cdot}b(p+q)\cdot b 
\frac{1}{2}(3x-7 x^2+12 x^3-20 x^4)\right.\nonumber\\
&&+(p_\mu+q_\mu)\left[(p\cdot b+q\cdot b)^2 (-4 x+3 x^2)+(l\cdot b)^2 
\frac{1}{2}(x+3 x^2 -6 x^4)\right]\nonumber\\
&&+\left[ \frac{2}{\epsilon_{\rm IR}}+\ln\frac{m^2}{4\pi\mu^2}
+\ln\left(1-\frac{l^2}{m^2}x (1-x)\right)+2\gamma\right]\left[
\frac{1}{2}l_\mu l{\cdot}b(p+q)\cdot b (-x+4 x^2-5 x^3\right.
\nonumber\\
&&\left.\left. +3 x^4)\right] 
+\frac{1}{4}(p_\mu+q_\mu)\left[(p\cdot b+q\cdot b)^2 
(2 x- x^2- x^3+x^4)
+(l\cdot b)^2 (x^2 -x^3+x^4)\right]\right\}\nonumber\\
&+&\frac{e^2}{2\pi^2}\int_0^1 dx \frac{\gamma_\mu}{[m^2-l^2 x(1-x)]^3}
\left\{\left[ \frac{2}{\epsilon_{\rm IR}}+\ln\frac{m^2}{4\pi\mu^2}
+\ln\left(1-\frac{l^2}{m^2}x (1-x)\right)+\gamma\right]\right.\nonumber\\
&&{\times}\frac{1}{4}(2m^2-l^2) x (x-1)
\left[(p\cdot b+q\cdot b)^2+(2 x-1)^2 (l\cdot b)^2\right]\nonumber\\
&&+(p\cdot b+q\cdot b)^2\left[-\frac{1}{2}m^2 (19x+7 x^2)
+\frac{3}{4}l^2 (5x-6 x^2)\right]\nonumber\\
&&\left.\left.+(l\cdot b)^2[\frac{1}{2}m^2 (9 x-7 x^2-12 x^4)+
\frac{1}{4}l^2 (-9 x+13 x^2-4 x^3+12 x^4)]\right\}\right)u(p),
\label{eq41}
\end{eqnarray} 
where we have used
\begin{eqnarray}
&&p_\mu (p\cdot b)^2+q_\mu (q\cdot b)^2=\frac{1}{2}\left\{
(p_\mu+q_\mu)\left[(p\cdot b)^2+(q\cdot b)^2\right]
+l_\mu l{\cdot}b (p+q)\cdot b\right\};\nonumber\\
&&p_\mu (q\cdot b)^2+q_\mu (p\cdot b)^2=\frac{1}{2}\left\{
(p_\mu+q_\mu)\left[(p\cdot b)^2+(q\cdot b)^2\right]
-l_\mu l{\cdot}b (p+q)\cdot b\right\};\nonumber\\
&& p\cdot b p_\mu+q\cdot b q_\mu =\frac{1}{2}[(p_\mu+q_\mu) (p+q)\cdot b
+l_\mu l\cdot b];\nonumber\\
&&p\cdot b q_\mu+q\cdot b p_\mu =\frac{1}{2}[(p_\mu+q_\mu) (p+q)\cdot b
-l_\mu l\cdot b];\nonumber\\
&& (p\cdot b)^2+(q\cdot b)^2=
\frac{1}{2}[ (p\cdot b+q\cdot b)^2+ (l\cdot b)^2];\nonumber\\
&&p\cdot b q\cdot b=\frac{1}{4}[ (p\cdot b+q\cdot b)^2- (l\cdot b)^2].
\label{eq42}
\end{eqnarray}
Eqs.\,(\ref{eq36})---(\ref{eq41}) give the contribution 
to the one-loop on-shell vertex to second order in $b$,
\begin{eqnarray}
&&\bar{u}(q)\Lambda^{(2)}_{\mu}(p,q)u(p)=\nonumber\\
&&\bar{u}(q)\left(\frac{e^2}{8\pi^2}\int_0^1 dx\frac{1}{m^2-l^2 x(1-x)}
\left[b^2\gamma_\mu (x^2-2 x-1)-2 b_\mu b\hspace{-2mm}/(x^2+2 x-1)\right]
\right.  \nonumber\\
&+&\frac{e^2}{4\pi^2}
\int_0^1 dx \frac{mb_\mu (p+q){\cdot}b}{[m^2-l^2 x(1-x)]^2}
\left\{7x^2-5 x+(9 x-3 x^2)
\left[\frac{1}{\epsilon_{\rm IR}}+\ln\frac{m^2}{4\pi\mu^2}
\right.\right.\nonumber\\
&&\left.\left.+\ln\left(1-\frac{l^2}{m^2}x (1-x)\right)+\gamma\right]
\right\}\nonumber\\
&+&\frac{e^2}{4\pi^2}\gamma_\mu \int_0^1 dx \frac{1}{[m^2-l^2 x(1-x)]^2}
\left\{-\frac{1}{2}[b\cdot (p+q)]^2 (x^2+6 x)\right.\nonumber\\
&&-\frac{1}{2}(l\cdot b)^2(x-4 x^2-2 x^3+4 x^4)- m^2 b^2 (1+x^2)
+\frac{1}{2}l^2 b^2 (x^2+x+1)\nonumber\\
&&+\left(\frac{1}{\epsilon_{\rm IR}}+\gamma\right)
\left[\frac{7}{4}x(p{\cdot}b+q{\cdot}b)^2
+\frac{1}{4}(x-2 x^2) (l\cdot b)^2
+m^2 b^2 (1-x)^2\right.\nonumber\\
&&\left. +\frac{1}{2}l^2 b^2 (-1+x-x^2)\right]
+\left[\ln\frac{m^2}{4\pi\mu^2}
+\ln\left(1-\frac{l^2}{m^2}x (1-x)\right)\right]
\left[\frac{9}{4}x (p\cdot b+q\cdot b)^2 
\right.\nonumber\\
&&\left.\left.+m^2 b^2 (1-x) (1-3 x)+\frac{1}{2}l^2 b^2 
(-1+x-x^2)+\frac{1}{4}(l\cdot b)^2 (2 x^2-x)\right]\right\}\nonumber\\
&+&\frac{e^2}{4\pi^2}\int_0^1 dx \frac{b\hspace{-2mm}/}{[m^2-l^2 x(1-x)]^2}
\left\{b{\cdot}(p+q)(p_\mu+q_\mu)\frac{1}{2}(x^2+4x)\right.\nonumber\\
&&+\frac{1}{2}l\cdot bl_\mu(-2 x+5 x^2-4 x^3+4 x^4)+m^2 b_\mu (6 x^2-10 x)
+\left[\frac{1}{\epsilon_{\rm IR}}+\ln\frac{m^2}{4\pi\mu^2}\right.\nonumber\\
&&\left.\left.+\ln\left(1-\frac{l^2}{m^2}x (1-x)\right)+\gamma\right]
\left[-\frac{5}{2} (p+q){\cdot}b (p_\mu+q_\mu)x+\frac{1}{2}
(x-2x^2)l\cdot bl_\mu\right]\right\}\nonumber\\
&+&\frac{e^2}{4\pi^2}\int_0^1 dx \frac{mb^2 (p_\mu+q_\mu)}
{[m^2-l^2 x(1-x)]^2}
\left\{-3 x+\left(\frac{x^2}{2}-x\right) \left(\frac{1}{\epsilon_{\rm IR}}+
\gamma\right)\right.\nonumber\\
&&\left.+\left[\ln\frac{m^2}{4\pi\mu^2}
+\ln\left(1-\frac{l^2}{m^2}x (1-x)\right)\right]
\left(-\frac{x^2}{2}-x\right) \right\}\nonumber\\
&+&\frac{e^2}{2\pi^2}\int_0^1 dx \frac{ml_\mu l{\cdot}b}{[m^2-l^2 x(1-x)]^3}
\left[\frac{1}{2}(p+q)\cdot b 
(3x+ x^2-4 x^3-12 x^4)\right.\nonumber\\
&&\left.+2mb\hspace{-2mm}/ x (1-x) (1-2 x)^2\right]\nonumber\\
&+&\frac{e^2}{2\pi^2}\int_0^1 dx \frac{m(p_\mu+q_\mu)}{[m^2-l^2 x(1-x)]^3}
 \left\{(p\cdot b+q\cdot b)^2 (-4 x+3 x^2)+\frac{(l\cdot b)^2}{2} 
(x+11 x^2 \right. \nonumber\\ 
&& -16x^3+2 x^4)+l^2b^2 x^2 (1-x) (7-2 x)
-9 m^2 b^2 x (1-x)\nonumber\\
&&+\left[ \frac{2}{\epsilon_{\rm IR}}+\ln\frac{m^2}{4\pi\mu^2}
+\ln\left(1-\frac{l^2}{m^2}x (1-x)\right)+2\gamma\right]
\left[\frac{1}{4}(p\cdot b+q\cdot b)^2 (2 x-x^2-x^3+x^4) \right.\nonumber\\
&&\left.\left.+\frac{1}{4}(l\cdot b)^2 (x^2 -x^3+x^4)+
\frac{1}{2}l^2 b^2 x^2 (1-x)-m^2 b^2 x (1-x)\right]\right\}\nonumber\\
&+&\frac{e^2}{2\pi^2}\int_0^1 dx \frac{\gamma_\mu}{[m^2-l^2 x(1-x)]^3}
\left\{\left[ \frac{2}{\epsilon_{\rm IR}}+\ln\frac{m^2}{4\pi\mu^2}
+\ln\left(1-\frac{l^2}{m^2}x (1-x)\right)+2 \gamma\right]\right.\nonumber\\
&&{\times}\frac{1}{4}(2m^2-l^2) x (x-1)
\left[(p\cdot b+q\cdot b)^2+(2 x-1)^2 (l\cdot b)^2\right]\nonumber\\
&&+(p\cdot b+q\cdot b)^2\left[-\frac{1}{2}m^2 (15x+11 x^2)
+\frac{3}{4}l^2 (5x- x^2)\right]\nonumber\\
&&+(l\cdot b)^2[\frac{1}{2}m^2 (17 x-31 x^2+32 x^3-28 x^4)+
\frac{1}{4}l^2 (l\cdot b)^2(-9 x+13 x^2-4 x^3+12 x^4)]\nonumber\\
&&\left.+m^2 b^2\left[4m^2 (3 x-7x^2+x^3)+l^2 (-2 x+6 x^2-x^3+2 x^4)\right]
\right\}\nonumber\\
&&+\frac{e^2}{2\pi^2}\int_0^1 dx\frac{mb_\mu (p+q){\cdot}b}
{[m^2-l^2 x (1-x)]^3}\left\{(8m^2+l^2)+
(4m^2-l^2)\left[\frac{2}{\epsilon_{\rm IR}}
+\ln\frac{m^2}{4\pi\mu^2}\right.\right.\nonumber\\
&&\left.\left.
+\ln\left(1-\frac{l^2}{m^2}x (1-x)\right)+2 \gamma\right]\right\}\nonumber\\
&+&\left.\frac{e^2}{\pi^2}\int_0^1 dx\frac{m^2 b\hspace{-2mm}/ x (1-x)}
{[m^2-l^2 x (1-x)]^2}\left[(2 l^2-12 m^2)b_\mu-2 l\cdot b l_\mu\right]
\right)u(p).
\label{eq43}
\end{eqnarray}

\subsection{On-shell Vertex Renormalization and Radiative Correction}

With the results given in Eqs.\,(\ref{eq30a}), (\ref{eq32a2}) 
and (\ref{eq43}), and employing the Gordon
identity (\ref{eqgor1}), we finally get the one-loop 
on-shell quantum vertex of  QED with
the Lorentz and $CPT$ violation term in the fermionic sector 
to the second order in $b_\mu$,
 \begin{eqnarray}
\Lambda_\mu (p,q,b)=\Lambda_\mu^{(0)}+ \Lambda_\mu^{(1)}+\Lambda_\mu^{(2)}.
\end{eqnarray}
The UV divergence only arises in  the conventional QED part. To be
consistent with the physical results of conventional QED, we define
the vertex renormalization and its radiative correction in the following way 
\begin{eqnarray}
\Lambda_{\mu}=\gamma_\mu\left(Z_1^{-1}-1\right)+Z_1^{-1}\Lambda^R_{\mu},
\end{eqnarray}
and the renormalization condition on the radiative correction part is
\begin{eqnarray}
\Lambda_{\mu}^R|_{q\hspace{-1.5mm}/=p\hspace{-1.5mm}/=m,\,l^2=0,\,b_\mu=0}=0.
\end{eqnarray}
Thus the vertex renormalization constant $Z_1$  is the same as the
conventional QED\cite{ref15},
\begin{eqnarray}
Z_1=1-\frac{e^2}{4\pi}\left(\frac{1}{\epsilon_{\rm UV}}-
\frac{2}{\epsilon_{\rm IR}}
+3\ln\frac{4\pi}{m^2}-4-3\gamma\right).
\end{eqnarray}
The radiative correction then consists of the conventional 
QED part and $\Lambda_{\mu}^{(1)}$, 
$\Lambda_{\mu}^{(2)}$, listed in (\ref{eq32a2}) and (\ref{eq43}),
\begin{eqnarray} 
\Gamma_{\mu}^R&=&\frac{\alpha}{4\pi}\gamma_\mu\int_0^1dx
\left\{2\left(\frac{1}{\epsilon_{\rm IR}}+\gamma\right)
-2\ln\frac{4\pi\mu^2}{m^2}+3-\ln\left(1-\frac{l^2}{m^2}x (1-x)\right)
\right.\nonumber\\
&&-\frac{2m^2-l^2}{m^2-l^2 x (1-x)}\left[\frac{1}{\epsilon_{\rm IR}}+\gamma
+\ln\frac{4\pi\mu^2}{m^2}+\ln \left(1-\frac{l^2}{m^2}x (1-x)\right)\right]
\nonumber\\
&&\left.+\frac{l^2 x (1-x)-6m^2 x}{m^2-l^2 x (1-x)}\right\}
+\frac{\alpha}{2\pi}i\sigma_{\mu\nu}ml^{\nu}\int_0^1dx
\frac{x}{m^2-l^2 x (1-x)}+\Lambda_{\mu}^{(1)}+\Lambda_{\mu}^{(2)}.
\end{eqnarray}
In the vicinity of $l^2=0$, we have the radiative correction
to the leading order of $l^2$,
\begin{eqnarray}
\Gamma_{\mu}^R&=&\frac{\alpha}{4\pi}\left\{\gamma_\mu\left[
\frac{2}{3}\left(\frac{1}{\epsilon_{\rm IR}}+\gamma
+\ln\frac{m^2}{4\pi\mu^2}\right)
-\frac{1}{2}\right]\frac{l^2}{m^2}+\frac{i\sigma_{\mu\nu}l^{\nu}}{m}
\left(1+\frac{1}{6}\frac{l^2}{m^2}\right)\right\}\nonumber\\
&+&\frac{\alpha}{m^2}\left\{(p+q)\cdot b\gamma_\mu\left[
-\frac{3}{4}-\frac{5}{8}\frac{l^2}{m^2}+
\left(\frac{1}{\epsilon_{\rm IR}}+\ln\frac{m^2}{4\pi\mu^2}+\gamma\right)
\left(-\frac{3}{2}+\frac{5}{12}\frac{l^2}{m^2}\right)\right]\right.
\nonumber\\
&+&(p_\mu+q_\mu) b\hspace{-2mm}/
\left[\frac{7}{4}+\frac{3}{8}\frac{l^2}{m^2}
-\frac{1}{3}\frac{l^2}{m^2}\left(\frac{1}{\epsilon_{\rm IR}}
+\ln\frac{m^2}{4\pi\mu^2}+\gamma\right)\right]
+m[b\hspace{-2mm}/,\gamma_\mu]\left[\frac{3}{8}
-\frac{5}{16}\frac{l^2}{m^2}\right.\nonumber\\
&&\left.\left.+\left(\frac{1}{\epsilon_{\rm IR}}
+\ln\frac{m^2}{4\pi\mu^2}+\gamma\right)
\left(\frac{3}{4}+\frac{1}{24}\frac{l^2}{m^2}\right)\right]
-\left(\frac{1}{4}+\frac{1}{24}\frac{l^2}{m^2}\right)m b_\mu\right\}\gamma_5
\nonumber\\
&+&\frac{\alpha}{\pi}\gamma_\mu\left\{-\frac{41}{6}\frac{b^2}{m^2}-
\frac{77}{6}\frac{(p\cdot b+q\cdot b)^2}{m^4}
+\frac{l^2}{m^2}\left[\frac{127}{120}\frac{b^2}{m^2}-
\frac{8}{3}\frac{(p\cdot b+q\cdot b)^2}{m^4}\right]
+\frac{5}{6}\frac{(l\cdot b)^2}{m^4}\right.\nonumber\\
&&+\left(\frac{1}{\epsilon_{\rm IR}}+\gamma\right)\left[
\frac{13}{24}\frac{(p\cdot b+q\cdot b)^2}{m^4}+\frac{1}{3}
\frac{b^2}{m^2}+\frac{l^2}{m^2}\left(\frac{31}{120}
\frac{(p\cdot b+q \cdot b)^2}{m^4}-\frac{19}{60}\frac{b^2}{m^2}\right)
\right.\nonumber\\
&&\left.-\frac{13}{120}\frac{(l\cdot b)^2}{m^4}\right]
+\ln\frac{m^2}{4\pi\mu^2}\left[\frac{23}{24}
\frac{(p\cdot b+q\cdot b)^2}{m^4}+
\frac{l^2}{m^2}\left(\frac{43}{120}\frac{(p\cdot b+q\cdot b)^2}{m^4}
-\frac{9}{20}\frac{b^2}{m^2}\right)\right.\nonumber\\
&&\left.\left.+\frac{1}{120}\frac{(l\cdot b)^2}{m^4}\right]\right\}
+\frac{\alpha}{\pi}\frac{b\hspace{-2mm}/}{m}
\left\{-\frac{34}{3}\frac{b_\mu}{m}+
\frac{(p+q)\cdot b (p_\mu+q_\mu)}{m^3}\left[\frac{7}{6}
-\frac{5}{4}\left(\frac{1}{\epsilon_{\rm IR}}+\ln\frac{m^2}{4\pi\mu^2}
+\gamma\right)\right]\right.\nonumber\\
&+&\left.\frac{l^2}{m^2}\left[-\frac{403}{60}\frac{b_\mu}{m} 
+\frac{(p+q)\cdot b (p_\mu+q_\mu)}{m^3}\left(\frac{71}{120}-
\frac{5}{12}\left(\frac{1}{\epsilon_{\rm IR}}+\ln\frac{m^2}{4\pi\mu^2}
+\gamma\right)\right)-\frac{29}{30}\frac{l\cdot bl_\mu}{m^3}\right]
\right\}\nonumber\\
&+&\frac{\alpha}{\pi}\frac{(p+q)\cdot b b_\mu}{m^3}
\left[\frac{5}{2}+\frac{37}{6}\left(\frac{1}{\epsilon_{\rm IR}}+\gamma\right)
+\frac{29}{6}\ln\frac{m^2}{4\pi\mu^2}\right]
+\frac{\alpha}{\pi}\frac{l_\mu l\cdot b}{m^3}
\left(-\frac{47}{30}\frac{(p+q)\cdot b}{m^3}\right)\nonumber\\
&+&\frac{\alpha}{\pi}\frac{p_\mu+q_\mu}{m}
\left\{-\frac{9}{2}\frac{b^2}{m^2}-2\frac{(p\cdot b+q\cdot b)^2}{m^4}
+\frac{l^2}{m^2}\left[-\frac{33}{40}\frac{b^2}{m^2}
-\frac{323}{280}\frac{(p\cdot b+q\cdot b)^2}{m^4}
\right]\right.\nonumber\\
&&+\frac{17}{30}\frac{(l\cdot b)^2}{m^4}+
\left(\frac{1}{\epsilon_{\rm IR}}+\gamma\right)\left[-\frac{1}{3}
\frac{b^2}{m^2}+\frac{37}{120}\frac{(p\cdot b+q\cdot b)^2}{m^4}+
\frac{l^2}{m^2}\left(
\frac{9}{56}\frac{(p\cdot b+q\cdot b)^2}{m^4}\right.\right.
\nonumber\\
&&\left.\left.-\frac{11}{60}\frac{b^2}{m^2}\right)+
\frac{17}{120}\frac{(l\cdot b)^2}{m^4}\right]
+\ln\frac{m^2}{4\pi\mu^2}\left[-\frac{4}{3}\frac{b^2}{m^2}+
\frac{37}{120}\frac{(p\cdot b+q\cdot b)^2}{m^4}\right.\nonumber\\
&&\left.\left.+
\frac{l^2}{m^2}\left(-\frac{1}{4}\frac{b^2}{m^2}+
\frac{9}{56}\frac{(p\cdot b+q\cdot b)^2}{m^4}\right)
+\frac{17}{120}\frac{(l\cdot b)^2}{m^4}\right]\right\}.
\label{eq54}
\end{eqnarray}

\section{Anomalous Magnetic Moment}

Applying the  Gordon identities (\ref{eqgor1}) and (\ref{eqgor2})
in the radiative corrections given in Eq.\,(\ref{eq54}),
 we see that in addition to
the one arising in the conventional QED,
there is a contribution to the anomalous
magnetic moment  from the terms
$mb^2(p_\mu+q_\mu)$, $m(p_\mu+q_\mu) (l\cdot b)^2$ and 
$m(p_\mu+q_\mu)l^2b^2$ in $\Lambda_{\mu}^{(2)}$. The 
part of the interaction Hamiltonian coming from 
the anomalous magnetic moment of a charged spinning particle 
with a slowing varying external magnetic field is thus
\begin{eqnarray}
\Delta H&=&e\left\{\frac{1}{4m}\frac{\alpha}{2\pi}+
\frac{b^2}{2m^3}\frac{\alpha}{\pi}\left[\frac{9}{2}
+\frac{1}{3}\left(\frac{1}{\epsilon_{\rm IR}}+\gamma\right)+
\frac{4}{3}\ln\frac{m^2}{4\pi\mu^2}\right]\right\}
\int d^3 x\bar{\psi}(x)\sigma_{\mu\nu}\psi (x)F^{\mu\nu}_c\nonumber\\
&-&e\frac{\alpha}{\pi}\frac{b^2}{2m^5}\left[\frac{33}{40}
+\frac{11}{60}\left(\frac{1}{\epsilon_{\rm IR}}+\gamma\right) 
+\frac{1}{4}\ln\frac{m^2}{4\pi\mu^2}\right]
\int\bar{\psi}(x)\sigma_{\mu\nu}\psi (x)\partial^2F^{\mu\nu}_c\nonumber\\
&+&e\frac{\alpha}{\pi}\frac{1}{2m^5}\left[\frac{17}{30}
+\frac{17}{120}\left(\frac{1}{\epsilon_{\rm IR}}+\gamma
+\ln\frac{m^2}{4\pi\mu^2}\right)\right]
\int\bar{\psi}(x)\sigma_{\mu\nu}\psi (x)(b\cdot\partial)^2F^{\mu\nu}_c(x),
\end{eqnarray}
where $F_{c\mu\nu}=\partial_\mu A_{c\nu}-\partial_\nu A_{c\mu}$ 
and $A_{c\mu}$ is the classical electromagnetic potential. Choosing
$F$ to be a constant magnetic field, $B_i=-B^i=\epsilon_{ijk}F^{jk}/2$,
and using $\sigma_{ij}=\epsilon_{ijk}\sigma^k$, we get the magnetic dipole
energy contributed by the anomalous magnetic moment
\begin{eqnarray}
-{\bf B}\cdot {\bf \mu}=-{\bf B}\cdot \left\{\frac{e}{2m} \frac{\alpha}{\pi}
\left[\frac{1}{2}+\frac{b^2}{m^2}\left(\frac{9}{2}
+\frac{1}{3}\left(\frac{1}{\epsilon_{\rm IR}}+\gamma\right)+
\frac{4}{3}\ln\frac{m^2}{4\pi\mu^2}\right)\right]2\int d^3x\bar{\psi}(x)
\frac{{\bf \sigma}}{2}\psi (x)\right\}.
\end{eqnarray}
Thus the modification to the gyromagnetic ration by the 
quantum correction is 
\begin{eqnarray}
a=\frac{1}{2}(g-2)=\frac{\alpha}{\pi}
\left[\frac{1}{2}+\frac{b^2}{m^2}\left(\frac{9}{2}
+\frac{1}{3}\left(\frac{1}{\epsilon_{\rm IR}}+\gamma\right)+
\frac{4}{3}\ln\frac{m^2}{4\pi\mu^2}\right)\right].
\label{eq57}
\end{eqnarray}

Comparing with the general result in conventional QED, one can see that
there arises additional contributions stemming from the Lorentz and 
$CPT$ violation. This result makes us embarrassed about the
physical validity of introducing a
Lorentz and $CPT$ violating term in the fermionic sector of QED and
hence the radiatively induced Lorentz and $CPT$ violation. The
anomalous magnetic moment is a measurable physical quantity
and it will yield an effective interaction Hamiltonian of QED. 
There is no mechanism to cancel the IR divergence in the anomalous 
magnetic moment terms.
In conventional QED, the anomalous magnetic moment is completely free
from IR divergence and hence gives a physically reasonable result.
Thus this seems to strongly suggest that the way of constructing a
QED model with Lorentz and $CPT$ violation by directly 
adding an explicit Lorentz and $CPT$ term breaking term
should be abandoned. In the following sections, we shall calculate
the Lamb shift, the prediction on which is another important 
achievement of QED, to see whether the IR divergence
it contains can be canceled like in the conventional 
case, i.e., whether it can be canceled by the IR divergence 
contributed from bremsstrahlung.

\section{Lamb shift}

\subsection{Radiative Correction to Classical Coulomb Potential}

Another well known achievement of QED is the precise correspondence
between the theoretical prediction and experimental measurement of Lamb shift.
Theoretically, the Lamb shift arises from the modification to the classical
Coulomb interaction of the radiative correction. The interaction of the
electron with the classical Coulomb potential produced from a nuclear 
of charge $-Ze$ is 
\begin{eqnarray}
V(r)=e\bar{\psi}(x)\gamma_0\psi (x)A_{\rm cl}^0(x)
=-Ze\frac{\psi^{\dagger}(x)\psi (x)}{4\pi r}, ~~~~r=|{\bf x}|,
\end{eqnarray}
and in momentum space it is written as
\begin{eqnarray}
V=Ze\frac{\bar{u}\gamma_0 u}{|{\bf l}|^2}
=Ze\frac{u^{\dagger}u}{|{\bf l}|^2}.
\end{eqnarray}
Thus to calculate the Lamb shift we need
to consider all possible radiative corrections to the tree level vertex
$\bar{\psi}(x)\gamma_\mu\psi (x) A^{\mu}$. As shown in Fig.\,1,
 this includes not only
the vertex radiative correction from the 1PI part, but also the 
self-energy in the fermionic external lines and the polarization 
tensor  in the external photon line. 
However, since in the on-shell renormalization
schemes, the quantum correction of the fermionic self-energy
takes the following form
 \begin{eqnarray}
\Sigma (p)=\Sigma (m)+(p\hspace{-2mm}/-m)B(m)+(p\hspace{-2mm}/-m)^2 
C(p,m).
\end{eqnarray}
The first term and the second one contribute to the mass 
renormalization and wave function renormalization of the electron,
and are canceled by the corresponding counterterms. The radiative
correction $\Sigma_R (p)$ of the fermionic self-energy is 
proportional to $(p\hspace{-2mm}/-m)^2$, thus the contribution to
the quantum vertex from the diagrams with the self-energy 
insertion in the fermionic external lines vanishes since
the amplitude is read as
\begin{eqnarray}
\bar{u}(q)\left[\Sigma_R(q)\frac{1}{q\hspace{-2mm}/-m}\gamma_\mu 
+\gamma_\mu \frac{1}{p\hspace{-2mm}/-m}\Sigma_R(p)\right] u(p). 
\end{eqnarray}
Therefore, we only consider the contribution  to the
Lamb shift from on-shell vertex correction and
the polarization tensor insertion in the external photon field, the
radiative correction to the classical interaction vertex, 
according to Fig.\,1, is thus read as 
\begin{eqnarray}
 e\bar{u}(q)\gamma_\mu u(p)\longrightarrow e\bar{u}(q)
\left[\gamma_\mu+\gamma^\nu D^{(1)}_{\nu\mu}(l)+\Lambda_\mu\right]u(p). 
\label{eq62}
\end{eqnarray}
The second term of (\ref{eq62}) comes from the insertion of 
vacuum polarization tensor, in the static case and Feynman 
gauge, $l^2=-{\bf l}^2$, the observed electric charge due to 
the screening of vacuum polarization is 
\begin{eqnarray}
\frac{e}{1+\omega_R ({\bf l},b)}
{\simeq}e\left[1+\left(\frac{1}{15}\frac{\alpha}{\pi}+
\frac{4}{15}\frac{\alpha}{\pi}\frac{b^2}{m^2}\right)\frac{{\bf l}^2}{m^2}
+\frac{4}{15}\frac{\alpha}{\pi}\frac{({\bf l}{\cdot}{\bf b})^2}{m^4}\right].
\end{eqnarray}
In configuration space of an infinitely heavy nucleus of charge
$-Ze$ located at the origin, the Coulomb potential is
 $A_{{\rm cl}\mu}=-g_{0\mu}Ze/(4\pi r)$. The modification of electric
charge screening to the classical Coulomb
potential  is thus
\begin{eqnarray}
&&\Delta V^{(1)}_{\rm eff}(r)=-\left[\left(\frac{1}{15}\frac{\alpha}{\pi}+
\frac{4}{15}\frac{\alpha}{\pi}\frac{b^2}{m^2}\right)\frac{{\bf \nabla}^2}{m^2}
+\frac{4}{15}\frac{\alpha}{\pi}\frac{({\bf \nabla}{\cdot}{\bf b})^2}{m^4}
\right]A_{c\mu}\nonumber\\
&&=-\left\{\frac{\alpha}{15\pi}\frac{Ze^2}{m^2}
\delta^{(3)}({\bf r})+\frac{4\alpha}{15\pi}\frac{Ze^2{\bf b}^2}{m^4}
\delta^{(3)}({\bf r})+\frac{4\alpha}{15\pi}\frac{1}{m^4}\frac{Ze^2}{4\pi r}
\left[\frac{{\bf b}^2}{r^2}-3\frac{({\bf b}{\cdot}{\bf r})^2}{r^4}\right]
\right\}\delta_{\mu 0}.
\label{eq64}
\end{eqnarray}

The effects from the on-shell vertex radiative correction are quite 
complicated. Eq.\,(\ref{eq54}) shows that only some of the terms
proportional to $\gamma_\mu$ and $p_\mu+q_\mu$ contribute to Lamb shift.
The corresponding contribution to the second term of (\ref{eq62}) is
\begin{eqnarray}
&&\frac{\alpha}{\pi}\bar{u}(q)\gamma_\mu  \left\{
\left[\frac{2}{3}\left(\frac{1}{\epsilon_{\rm IR}}+\ln\frac{m^2}{4\pi\mu^2}+
\gamma\right)-\frac{1}{2}\right]\frac{l^2}{4m^2}
-\frac{581}{60}\frac{b^2}{m^2}-\frac{71}{120}\frac{l^2b^2}{m^4}
\right.\nonumber\\
&&+\frac{59}{30}\frac{(l\cdot b)^2}{m^4}
+\left(\frac{1}{\epsilon_{\rm IR}}+\gamma\right)
\left[-\frac{1}{3}\frac{b^2}{m^2}
-\frac{41}{60}\frac{l^2b^2}{m^4}+\frac{7}{40}\frac{(l\cdot b)^2}{m^4}\right]
\nonumber\\
&&\left.+\ln\frac{m^2}{4\pi\mu^2}\left[-\frac{8}{3}\frac{b^2}{m^2}
-\frac{19}{20}\frac{l^2b^2}{m^4}+\frac{7}{24}\frac{(l\cdot b)^2}{m^4}\right]
\right\}u(p)\nonumber\\
&&+\frac{\alpha}{\pi}\bar{u}(q)
\frac{i\sigma_{\mu\nu}l^{\nu}}{m}\left\{\frac{1}{4}\left(1+\frac{1}{6}
\frac{l^2}{m^2}\right)
+\frac{9}{2}\frac{b^2}{m^2}+\frac{33}{40}\frac{l^2b^2}{m^4}-
\frac{17}{30}\frac{(l\cdot b)^2}{m^4}
+\left(\frac{1}{\epsilon_{\rm IR}}+\gamma\right)\right.\nonumber\\
&&\left.{\times}\left[\frac{1}{3}\frac{b^2}{m^2}
+\frac{11}{60}\frac{l^2b^2}{m^4}
-\frac{17}{120}\frac{(l\cdot b)^2}{m^4}\right]
+\ln\frac{m^2}{4\pi\mu^2}\left[\frac{4}{3}\frac{b^2}{m^2}
+\frac{1}{4}\frac{l^2b^2}{m^4}
-\frac{17}{120}\frac{(l\cdot b)^2}{m^4}\right]\right\}u(p).
\label{eq65}
\end{eqnarray}
In the static case, $l^2=-{\bf l}^2$, with the replacement
${\bf l}=-i{\bf \nabla}$, Eq.\,(\ref{eq65}) leads to the following 
correction to the Coulomb potential,
\begin{eqnarray}
\Delta V^{(2)}_{\rm eff}&=&\Delta V^{(2)\prime}_{\rm eff}
+\Delta V^{(2)\prime\prime}_{\rm eff}, \nonumber\\
\Delta V^{(2)\prime}_{\rm eff}&=&\frac{Z\alpha^2}{\pi}\frac{b^2}{m^2}\left[
\frac{581}{60}+\frac{1}{3}\left(\frac{1}{\epsilon_{\rm IR}}+\gamma\right)+
\frac{8}{3}\ln\frac{m^2}{4\pi\mu^2}\right]\frac{1}{r}
+4Z{\alpha}^2\left\{\frac{1}{6m^2}\left(\frac{1}{\epsilon_{\rm IR}}
+\gamma \right.\right.\nonumber\\
&&+\left.\left.\ln\frac{m^2}{4\pi\mu^2}\right)- \frac{1}{8m^2}
-\frac{b^2}{m^4}\left[\frac{71}{120}+\frac{41}{60}
 \left(\frac{1}{\epsilon_{\rm IR}}+\gamma\right)+\frac{19}{20}
\ln\frac{m^2}{4\pi\mu^2}\right]\right\}\delta^{(3)}({\bf x})
\nonumber\\
&&+\frac{Z\alpha^2}{\pi}\frac{1}{m^4}\left[\frac{59}{30}+\frac{7}{40}
 \left(\frac{1}{\epsilon_{\rm IR}}
+\gamma\right)+\frac{7}{24}\ln\frac{m^2}{4\pi\mu^2}\right]
\frac{3 ({\bf x}\cdot {\bf b})^2-r^2 {\bf b}^2}{r^5};\nonumber\\
\Delta V^{(2)\prime\prime}_{\rm eff}&=&
\frac{ie\alpha}{\pi m}
\left\{\frac{1}{4}\left(1+\frac{1}{6}\frac{{\bf \nabla}^2}{m^2}
\right)+\frac{b^2}{m^2}\left(\frac{9}{2}
+\frac{33}{40}\frac{{\bf \nabla}^2}{m^2}\right)+
\frac{17}{30}\frac{({\bf b}\cdot {\bf \nabla})^2}{m^4}\right.\nonumber\\
&&+\left(\frac{1}{\epsilon_{\rm IR}}+\gamma\right)\left[
\frac{b^2}{m^2}\left(\frac{1}{3}
+\frac{11}{60}\frac{{\bf \nabla}^2}{m^2}\right)+
\frac{17}{120}\frac{({\bf b}\cdot {\bf \nabla})^2}{m^4}\right]\nonumber\\
&&+\left.\ln\frac{m^2}{4\pi\mu^2}\left[\frac{b^2}{m^2}\left(\frac{4}{3}
+\frac{1}{4}\frac{{\bf \nabla}^2}{m^2}\right)+
\frac{17}{120}\frac{({\bf b}\cdot {\bf \nabla})^2}{m^4}\right]\right\}
\left[{\bf \gamma}\cdot {\bf E}(r)\right],
\end{eqnarray}
where ${\bf E}=-{\bf \nabla} A_0=-Ze/(4\pi){\bf x}/{r^3}$ is the
static Coulomb electric field. The term with electric field
implies that the anomalous magnetic moment induces an electric
dipole moment for a moving electron.
Writing the electron spinor wave function
in two-component form,
$\psi=\left(\varphi, \chi\right)^T$, 
$\psi^{\dagger}=\left(\varphi^{\dagger},\chi^{\dagger}\right)$,
and using the Pauli's non-relativistic approximation\cite{bjoken},
the large two-component $\varphi$ and the the small one $\chi$ being
related by
\begin{eqnarray}
\chi (r)=-i\frac{{\bf \sigma}{\cdot}{\bf \nabla}}{2m}{\varphi}(r),
\end{eqnarray}
the anomalous magnetic part can be re-written as the form  with the 
spin-orbit angular momentum interaction,
\begin{eqnarray}
\Delta V^{(2)\prime\prime}_{\rm eff}&=& \frac{Z\alpha^2}{2\pi m^2}
\left\{\frac{1}{4}+\frac{b^2}{m^2}\left[\frac{9}{2}+\frac{1}{3}\left(
1+\frac{1}{\epsilon_{\rm IR}}+\gamma\right)+\frac{4}{3}
\ln\frac{m^2}{4\pi\mu^2}\right]\right\}
\left[4\pi\delta^{(3)}({\bf x})+4\frac{{\bf S}\cdot {\bf L}}{r^3}\right]
\nonumber\\
&+&\frac{2Z\alpha^2}{m^4}\left\{\frac{1}{24}+\frac{b^2}{m^2}
\left[\frac{33}{40}+\frac{11}{60}\left(\frac{1}{\epsilon_{\rm IR}}+\gamma\right)
+\frac{1}{4}\ln\frac{m^2}{4\pi\mu^2}\right]\right\}\nonumber\\
&&\times \left\{{\bf \nabla}^2
\delta^{(3)}({\bf x})-2i{\bf S}\cdot\left[{\bf \nabla}\delta^{(3)}({\bf x})
\times {\bf \nabla}\right]\right\}\nonumber\\
&+&\frac{Z\alpha^2}{2\pi m^6}\left\{
4\pi\left({\bf b}\cdot{\bf \nabla}\right)^2\delta^{(3)}({\bf x})
+12\left[\frac{5({\bf b}\cdot{\bf x})^2}{r^7}-\frac{b^2}{r^5}\right]
{\bf L}{\cdot}{\bf S}\right.\nonumber\\
&&\left.-24\frac{i({\bf x}{\cdot}{\bf b}){\bf b}\cdot
({\bf S}\times{\bf \nabla})}{r^5}\right\} \left[\frac{17}{30}+\frac{17}{120}
\left(\frac{1}{\epsilon_{\rm IR}}+\gamma+\ln\frac{m^2}{4\pi\mu^2}\right)\right].
\label{eq68}
\end{eqnarray}
where ${\bf S}={\bf \sigma}/2$, ${\bf L}=i{\bf x}{\times}{\bf \nabla}$.
Note that the spin-orbit interaction arises from the anomalous
magnetic moment. In deriving (\ref{eq68}) we have used 
\begin{eqnarray}
&&\psi^{\dagger}\frac{{\bf \gamma}\cdot {\bf x}}{r^3}\psi
=\varphi^{\dagger}\frac{{\bf \sigma}{\cdot}{\bf x}}{r^3}{\chi}-
{\chi}^{\dagger}\frac{{\bf \sigma}{\cdot}{\bf x}}{r^3}\varphi
=\frac{1}{2im}\varphi^{\dagger}[\frac{{\bf \sigma}\cdot {\bf x}}{r^3},
{\bf \sigma}\cdot {\bf \nabla}]\varphi,\nonumber\\
&&[{\bf \sigma}\cdot{\bf A}, {\bf \sigma}\cdot{\bf B}]
={\bf A}\cdot {\bf B}-{\bf B}\cdot {\bf A}+i{\bf \sigma}
\left({\bf A}\times {\bf B}+{\bf B}\times {\bf A}\right).
\end{eqnarray}
 
\subsection{Energy Level Shift of Hydrogen-Like Atom due to Radiative
Correction}

In the following we consider the energy-level shift in 
the hydrogen-like atom due to the radiative correction to the
classical Coulomb potential. 
Let us first look at the displacement of the energy-level
due to the screening of charge implied by the vacuum polarization.
First, the $\delta$-potential of Eq.\,(\ref{eq64}) 
will lead to a displacement of the $s$-state energy level, 
\begin{eqnarray} 
\delta_1^{\prime} E_{n,l}&=&-\left(\frac{Z\alpha}{15\pi}\frac{e^2}{m^2}+
\frac{4Z\alpha}{15\pi}\frac{e^2{\bf b}^2}{m^4}\right)\int d^3r
\psi^{\star}_{n,l}({\bf r})\delta^{(3)}({\bf r})\psi_{n,l}({\bf r})
 \nonumber\\
&=&-\left(\frac{Z\alpha}{15\pi}\frac{e^2}{m^2}+
\frac{4Z\alpha}{15\pi}\frac{e^2{\bf b}^2}{m^4}\right)\delta_{l,0}
|\psi_{n,0}|^2
=-\left(1+\frac{4{\bf b}^2}{m^2}\right)\frac{4}{15\pi}
\frac{Z^4\alpha^5}{n^3}m\delta_{l,0},
\end{eqnarray}
it is $1+4{\bf b}^2/m^2$ times of the corresponding displacement
in the conventional QED.

The more interesting effect comes from the quadrupole-like part
proportional to ${\bf b^2}/r^3-3 ({\bf b}{\cdot}{\bf r})^2)/r^5$.
In the lowest order approximation, the contribution of this potential
to the energy-level of hydrogen-like atom is
\begin{eqnarray}
\delta_1^{\prime\prime} E_{n,l}
=-\frac{4Z\alpha}{15\pi}\frac{1}{m^4}\frac{e^2}{4\pi}
\int d^3r\psi^{\star}_{n,l}({\bf r})\left[\frac{{\bf b}^2}{r^3}
-3\frac{({\bf b}{\cdot}{\bf r})^2}{r^5}\right]\psi_{n,l}({\bf r}).
\end{eqnarray}
For convenience, taking ${\bf b}$ in the direction of $z$-axis,
 we then have
\begin{eqnarray}
\delta_1^{\prime\prime} E_{n,l}&=&-\frac{4}{15\pi}
\frac{{\bf b}^2Z\alpha^2}{m^4}
\langle n l |\frac{1}{r^3}|n l\rangle +
\frac{4}{5\pi} \frac{{\bf b}^2Z\alpha^2}{m^4} 
\langle n l |\frac{1}{r^3}|n l\rangle 
\langle l M |\cos^2\theta |l M\rangle \nonumber\\
&=&\frac{{\bf b}^2Z^4\alpha^5}{m\pi}\frac{8}{n^3(2 l+1)[(2 l+1)^2-1]}
\left\{-\frac{4}{15}+\frac{4}{5}\left[ 
\frac{(l+M)(l-M)}{(2l+1) (2l-1)}\right.\right.\nonumber\\
&&\left.\left.+\frac{(l+M+1)(l-M+1)}{(2l+1) (2l+3)}\right]\right\},
\label{eq27}
\end{eqnarray}
where we have used the orthogonality and recurrence relations 
of spherical function
\begin{eqnarray}
\langle l M |l^{\prime} M^{\prime}\rangle &=&
\int_0^\pi d\theta \int_0^{2\pi}d\varphi Y_{lM}^{\star}(\theta,\varphi)
Y_{l^{\prime}M^{\prime}}(\theta,\varphi)\sin\theta=\delta_{ll^{\prime}}
\delta_{MM^{\prime}};\nonumber\\
\cos\theta Y_{lM}&=&\sqrt{\frac{(l+M)(l-M)}{(2l+1) (2l-1)}}Y_{l-1,M}+
\sqrt{\frac{(l+M+1)(l-M+1)}{(2l+1) (2l+3)}}Y_{l+1,M};\nonumber\\
\langle nl|\frac{1}{r^3}|n l\rangle &=& \left\{\begin{array}{cc}
{8(mZ\alpha)^3}/\{(2l+1)n^3[(2l+1)^2-1]\}, & l>0\\ \infty &l=0\end{array}\right..
\label{eq74}
\end{eqnarray}
The energy level shift due to the screening of electric charge 
is thus
\begin{eqnarray}
\delta_1 E_{nl}&=&\delta_1^{\prime} E_{n,l}+\delta_1^{\prime\prime} E_{n,l}
\nonumber\\
&=&-\frac{4}{15}\frac{Z^4\alpha^5}{\pi n^3}m \delta_{l0}-
\frac{16}{15}\frac{{\bf b}^2}{m}\frac{Z^4\alpha^5}{\pi n^3}\left\{\delta_{l0}
+\frac{2}{(2l+1)[(2l+1)^2-1]}\right.\nonumber\\
&&{\times}\left. \left[1-3\left(\frac{(l+M)(l-M)}{(2l+1) (2l-1)}
+\frac{(l+M+1)(l-M+1)}{(2l+1) (2l+3)}\right)\right]\right\},
\label{eq74a}
\end{eqnarray}
and
\begin{eqnarray}
\delta_1E_{n0}=-\frac{4}{15}\frac{Z^4\alpha^5}{\pi n^3} 
m \left(1+\frac{4b^2}{m^2}\right).
\label{eq74b}
\end{eqnarray}

The modification of energy level from the vertex correction can be calculated
in a similar way. Since $b^2{\ll}m^2$, so we only consider the correction
to the order $b^2/m^4$. Using Eq.\,(\ref{eq74}) and the following formula
\begin{eqnarray}
&&\langle nl|\frac{1}{r}|nl\rangle =\frac{me^2}{n^2},~~
\langle jlM|{\bf S}\cdot {L}|jlM\rangle =
\frac{1}{2}\left[j(j+1)-l(l+1)-\frac{3}{4}\right], ~~j=l\pm\frac{1}{2},
\end{eqnarray}
we get the shift of energy-level in hydrogen-like atom, 
\begin{eqnarray}
\delta_2 E_{nlj}&=&\int d^3{\bf x}\psi_{nlj}^{\dagger}({\bf x})
\Delta V_{\rm eff}^{(2)}\psi_{nlj}^{\dagger}({\bf x})\nonumber\\
&=&\frac{4Z\alpha^3}{n^2}\frac{b^2}{m}
\left[\frac{581}{60}+\frac{1}{3}\left(\frac{1}{\epsilon_{\rm IR}}+\gamma\right)
+\frac{8}{3}\ln\frac{m^2}{4\pi\mu^2}\right]
+\frac{4Z^4\alpha^5}{\pi n^3}m\delta_{l0}\left\{\frac{1}{6}\left(
\frac{1}{\epsilon_{\rm IR}}+\gamma\right.\right.\nonumber\\
&&\left.\left.+\ln\frac{m^2}{4\pi\mu^2}\right)
-\frac{1}{8}-\frac{b^2}{m^2}\left[\frac{71}{120}+
\frac{41}{60}\left(\frac{1}{\epsilon_{\rm IR}}+\gamma\right)+
\frac{19}{20}\ln\frac{m^2}{4\pi\mu^2}\right]\right\} \nonumber\\
&+&\frac{8Z^4\alpha^5}{\pi n^3}\frac{b^2}{m}
\left[\frac{59}{30}+\frac{7}{40}\left(\frac{1}{\epsilon_{\rm IR}}+\gamma\right)
+\frac{7}{24}\ln\frac{m^2}{4\pi\mu^2}\right]
\frac{1}{(2l+1)[(2l+1)^2-1]}\nonumber\\
&&\times \left[\frac{3(l+M)(l-M)}{(2l-1) (2l+1)}+\frac{3(l+M+1)(l-M+1)}
{(2l+1) (2l+3)}-1\right]
+\frac{2Z^4\alpha^5}{\pi n^3}m\left\{\frac{1}{4}\right.\nonumber\\
&&\left.+\frac{b^2}{m^2}\left[\frac{9}{2}
+\frac{1}{3}\left(\frac{1}{\epsilon_{\rm IR}}+\gamma\right)+\frac{4}{3}
\ln\frac{m^2}{4\pi\mu^2}\right]\right\}\left[\delta_{l0}+
\frac{4j(j+1)-4l(l+1)-3}{(2l+1)[(2l+1)^2-1]}\right],
\label{eq76}
\end{eqnarray}
 and
\begin{eqnarray}
\delta_2E_{n0j}&=&\frac{4Z\alpha^3}{n^2}\frac{b^2}{m}
\left[\frac{581}{60}+\frac{1}{3}\left(\frac{1}{\epsilon_{\rm IR}}+\gamma\right)
+\frac{8}{3}\ln\frac{m^2}{4\pi\mu^2}\right]\nonumber\\
&+&\frac{2Z^4\alpha^5}{\pi n^3}m
\left\{\frac{1}{3}\left(\frac{1}{\epsilon_{\rm IR}}+\gamma
+\ln\frac{m^2}{4\pi\mu^2}\right)\right.\nonumber\\
&&+\left.\frac{b^2}{m^2}\left[\frac{199}{60}
-\frac{31}{30}\left(\frac{1}{\epsilon_{\rm IR}}+\gamma\right)
-\frac{17}{30}\ln\frac{m^2}{4\pi\mu^2}\right]\right\}.
\label{eq76a}
\end{eqnarray}

The total correction on the energy-level of hydrogen-like atom
is given by (\ref{eq74}) and (\ref{eq76}).

\subsection{Lamb Shift}

From (\ref{eq74}) and (\ref{eq76}), 
one can easily calculate the energy level splitting between
the states $2S_{1/2}$ and $2P_{1/2}$ of the hydrogen atom. It is well know that
in Dirac's relativistic electron theory, these two states have the same energy
level. The splitting due to the radiative correction of QED leads to the Lamb
shift. We have 
\begin{eqnarray}
E_{2S_{1/2}}-E_{2P_{1/2}}&=&\delta E_{2S_{1/2}}-\delta E_{2P_{1/2}}
=\frac{m\alpha^5}{4\pi}\left\{-\frac{1}{20}+\frac{1}{3}\left(
\frac{1}{\epsilon_{\rm IR}}+\gamma+\ln\frac{m^2}{4\pi\mu^2}\right)\right.
\nonumber\\
&+&\frac{b^2}{m}\left[\left(\frac{257}{60}-
\frac{17(6M^2-4)}{300}\right)
-\left(\frac{83}{90}+\frac{7(6M^2-4)}{50}\right)
\left(\frac{1}{\epsilon_{\rm IR}}+\gamma\right) \right.\nonumber\\
&-&\left.\left.
\left(\frac{11}{90}+\frac{7(6M^2-4)}{30}\right)
\ln\frac{m^2}{4\pi\mu^2}\right]\right\}, ~~~M=\pm 1, 0.
\label{eq79}
\end{eqnarray}
This shows that in addition to the usual Lamb shift in 
QED, there arises a contribution from the $CPT$-odd
fermionic term in the action. It is remarkable that in the $b$-dependent
part, the Lamb shift has a dependence on the magnetic quantum number $M$,
which means that the Lambs shift itself also has a hyperfine structure.  
The cancellation or not of the IR singularity will be discussed
in the following section.

\section{ Non-Cancellation of IR Divergence in Lamb Shift}

The Lamb shift given in Eq.\,(80)  cannot be compared to
the experimental measurement, since it contains 
IR divergent terms. Like in the conventional QED, we now consider
the contribution to the form factors of the vertex correction
from the bremsstrahlung processes and hope that the IR
divergence can be canceled. Unfortunately, we find  that 
there exist several serious problems which make it impossible
for the accomplishment of IR divergence cancellation. This appears to
put further in doubt the legitimacy of introducing the term 
$\bar{\psi}b\hspace{-2mm}/\gamma_5\psi$.

Since the cancellation of IR divergences is expected to occur at the level
of physical cross sections, we consider the scattering of 
an electron  by  the static Coulomb potential
produced from a nucleus with charge $-Ze$,
\begin{eqnarray}
A_0({\bf x})=A_{\mu}({\bf x})
\delta_{\mu 0}=-\frac{Ze}{|{\bf x}|}\delta_{\mu 0}
=-\delta_{\mu 0} Ze\int \frac{d^3{\bf k}}{(2\pi)^3}e^{-i{\bf k}\cdot{\bf x}}
\frac{1}{{\bf k}^2}.
\label{eq8a1}
\end{eqnarray} 
The corresponding tree-level (Born) differential cross section is
\begin{eqnarray}  
d\sigma_0&=&\frac{E_q}{|{\bf q}|}2\pi\delta (E_q-E_p)\frac{d^3q}{(2\pi)^3}
\frac{Z^2e^4}{|{\bf l}|^4}\frac{m^2}{E_pE_q}|M^{(0)}|^2,\nonumber\\
|M^{(0)}|^2&=&\frac{1}{2} 
\sum |\bar{u}(q)\gamma_0u(p)|^2.
\label{eq8a24}
\end{eqnarray}
Here and in the following the sum is over both  the final and initial 
spin polarization states of the electron, the factor $1/2$ coming from 
averaging over the spin polarization of the initial electron. 
The contribution to the cross section from the 
 one-loop quantum corrected vertex is 
\begin{eqnarray}
d\sigma_1=\frac{E_q}{|{\bf q}|}2\pi\delta (E_q-E_p)\frac{d^3q}{(2\pi)^3}
\frac{Z^2e^4}{|{\bf l}|^4}\frac{m^2}{E_pE_q}|M^{(1)}|^2,
\label{eq8a25}
\end{eqnarray}
where
\begin{eqnarray}
|M^{(1)}|^2&=&\frac{1}{2} 
\sum |\bar{u}(q)\delta_{\mu 0}\left[\gamma_\mu+\Lambda^{(1)}_\mu\right]u(p)|^2;
\nonumber\\
\bar{u}(q)\Lambda^{(1)}_\mu (p,q,b)u(p) 
&=&\gamma_\mu F_1+\frac{i\sigma_{\mu\nu}l^{\nu}}{m}F_2
+\gamma_\mu\gamma_5F_3+\frac{(p_\mu+q_\mu)b\hspace{-2mm}/\gamma_5}{m}F_4
\nonumber\\
&&+\frac{\epsilon_{\mu\nu\lambda\rho}\sigma^{\lambda\rho}b^{\nu}}{m}F_5
+\frac{b_\mu}{m}\gamma_5 F_6
+\frac{b\hspace{-2mm}/ b_\mu}{m^2}F_7
+\frac{b\hspace{-2mm}/ (p_\mu+q_\mu)(p+q)\cdot b}{m^3}F_8
\nonumber\\
&&+\frac{l\cdot b(p+q)\cdot bl_\mu b\hspace{-2mm}/}{m^6} F_9
+\frac{(p+q)\cdot b b_\mu}{m^3}F_{10}+\frac{l_\mu l\cdot b (p+q)\cdot b}
{m^5} F_{11},
\label{eq8a26}
\end{eqnarray}
$F_i{\equiv}F_i[l^2, b^2, (p+q)\cdot b]$ ($i=1,{\cdots},11$) near
$l^2=0$ can be read out from Eq.\,(55). Eqs.\,(\ref{eq8a24}),
(\ref{eq8a25}) and (\ref{eq8a26}) shows that due to
the Lorentz and $CPT$ violation term, 
$\bar{\psi}b\hspace{-2mm}/\gamma_5\psi$,
the radiative correction makes the tensor structure
of quantum vertex become much more complicated than that
in  conventional QED, in which
\begin{eqnarray}
|M^{(1)}|^2=|M^{(0)}|^2\left[1+F_1(l^2)\right]^2
{\simeq}|M^{(0)}|^2\left[1+2F_1(l^2)\right]
\label{eq5a}
\end{eqnarray}
Since this is not valid any longer, there does not exist the following relation
between the classical and quantum differential cross section, 
\begin{eqnarray}
d\sigma_1=[1+2F_1(l^2)] d\sigma_0. 
\label{eq6a}
\end{eqnarray}
This is actually the first obstacle for the IR divergence cancellation.

 In the following we consider the 
bremsstrahlung process:  photon emission from an electron when it
is scattered by the static Coulomb potential (\ref{eq8a1}).
As in conventional QED, the scattering matrix is
\begin{eqnarray}
S=\frac{iZe^3}{V^{3/2}}2\pi \delta (E_q+E_k-E_p)
\frac{m}{\sqrt{E_qE_p}}\frac{1}{|{\bf k}|^2}\epsilon_{\nu}^{(\lambda)}(k)
M^{\nu}_{1\gamma},
\end{eqnarray}
where $E_q$ and $E_p$ are the energy of the electron before and after
being scattered by the nucleus; $E_k$ is the energy of emitted photon
and $\epsilon_{\mu}^{(\lambda)}(k)$, $\lambda=1,2$ are the photon polarization 
vectors, which satisfy the orthogonality and the completeness relations
\begin{eqnarray}
g^{\mu\nu}\epsilon_{\mu}^{(\lambda)}(k)\epsilon_{\nu}^{(\sigma)}(k)
=\delta^{\lambda\sigma}, 
~~\sum_{\lambda=1,2}\epsilon_{\mu}^{(\lambda)}(k)\epsilon_{\nu}^{(\lambda)}(k)
=-\left(g_{\mu\nu}-\frac{k_{\mu}k_{\nu}}{k^2}\right);
\end{eqnarray}
$V$ is the normalization volume for the initial electron; 
$M_{(1\gamma)\nu}$ is the
matrix element,
\begin{eqnarray}
M_{(1\gamma)\nu}=\bar{u}(q)\left(\gamma_\nu
\frac{1}{k\hspace{-2mm}/+q\hspace{-2mm}/-m-b\hspace{-2mm}/\gamma_5}\gamma_0
+\gamma_0\frac{1}{p\hspace{-2mm}/-k\hspace{-2mm}/-m-b\hspace{-2mm}/\gamma_5}
\gamma_\nu\right)u(p),
\end{eqnarray}
the subscript $1\gamma$ denoting that we only consider the process of
 single photon emission. The corresponding 
differential cross section is given by 
the amplitude, $|S|^2$, per incoming electron flux
and time and  summing over the final states of the electrons and photons,
\begin{eqnarray}
d\sigma_{(1\gamma)} 
=\frac{Z^2e^6}{|{\bf v}|}\frac{m}{E_p}\frac{d^3q}{(2\pi)^3}
\frac{d^3k}{(2\pi)^3}\delta (E_q+E_k-E_p)\frac{m}{E_q}\frac{1}{2E_k}
\sum_{\lambda=1,2}|\epsilon_{\nu}^{(\lambda)}(k)M_{(1\gamma)}^\nu|^2,
\end{eqnarray}
In the above, ${\bf v}$ is  the velocity of the incoming electron, 
$|{\bf v}|=|{\bf p}|/m$.

We now evaluate $|\epsilon_{\nu}^{(\lambda)}(k)M^{\nu}|^2$
in the limit $k_\mu{\rightarrow}0$. Employing Eq.\,(10), we get the expansion
to the second order of $b$,
\begin{eqnarray}
M_{(1\gamma)\nu}&=&\bar{u}(q)\left[\gamma_\nu\left(
\frac{1}{k\hspace{-2mm}/+q\hspace{-2mm}/-m}+
\frac{1}{k\hspace{-2mm}/+q\hspace{-2mm}/-m}b\hspace{-2mm}/\gamma_5
\frac{1}{k\hspace{-2mm}/+q\hspace{-2mm}/-m}\right.\right.\nonumber\\
&&\left.+\frac{1}{k\hspace{-2mm}/
+q\hspace{-2mm}/-m}b\hspace{-2mm}/\gamma_5
\frac{1}{k\hspace{-2mm}/+q\hspace{-2mm}/-m}b\hspace{-2mm}/\gamma_5
\frac{1}{k\hspace{-2mm}/+q\hspace{-2mm}/-m}\right)\gamma_0
\nonumber\\
&&+\gamma_0\left(
\frac{1}{p\hspace{-2mm}/-k\hspace{-2mm}/-m}+
\frac{1}{p\hspace{-2mm}/-k\hspace{-2mm}/-m}b\hspace{-2mm}/\gamma_5
\frac{1}{p\hspace{-2mm}/-k\hspace{-2mm}/-m}\right.\nonumber\\
&& \left.\left.+\frac{1}{p\hspace{-2mm}/-k\hspace{-2mm}/-m}
b\hspace{-2mm}/\gamma_5
\frac{1}{p\hspace{-2mm}/-k\hspace{-2mm}/-m}b\hspace{-2mm}/\gamma_5
\frac{1}{p\hspace{-2mm}/-k\hspace{-2mm}/-m}\right)\gamma_0
+{\cdots}\right]u(p).
\label{eqa6}
\end{eqnarray}
The $b^0$ term in the soft photon limit (i.e.$k_\mu\rightarrow 0$) 
is  identical to that obtained in conventional QED,
\begin{eqnarray}
M_{\nu}(b^0)\stackrel{k\rightarrow 0}{\simeq}
\bar{u}(q)\gamma_0\left(\frac{q_\nu}{q\cdot k}-\frac{p_\nu}{p\cdot k}
\right)u(p)
\label{eqa7}.
\end{eqnarray}
As for the $b^1$ term, it is 
\begin{eqnarray}
M_\nu (b^1)&=&\bar{u}(q)\left[
\frac{\gamma_\nu (k\hspace{-2mm}/+q\hspace{-2mm}/+m)b\hspace{-2mm}/
(k\hspace{-2mm}/+q\hspace{-2mm}/-m)\gamma_0\gamma_5}{4(q\cdot k)^2}
+\frac{\gamma_5\gamma_0 (p\hspace{-2mm}/-k\hspace{-2mm}/-m)b\hspace{-2mm}/
(p\hspace{-2mm}/-k\hspace{-2mm}/+m)\gamma_\nu}{4(p\cdot k)^2}\right]
u(p)\nonumber\\
&\stackrel{k\rightarrow 0}{\simeq}&
\bar{u}(q)\left[\frac{2 q_\nu (k+q)\cdot b-2 q_\nu m b\hspace{-2mm}/
-q\cdot b k\hspace{-2mm}/\gamma_\nu -k\cdot q\gamma_\nu  
b\hspace{-2mm}/+m k\hspace{-2mm}/\gamma_\nu b\hspace{-2mm}/}{2(q\cdot k)^2}
\gamma_0\gamma_5\right.\nonumber\\
&&\left.+\gamma_5\gamma_0
\frac{2 p_\nu (p-k)\cdot b-2m p_\nu  b\hspace{-2mm}/
+p\cdot b\gamma_\nu k\hspace{-2mm}/+k\cdot p b\hspace{-2mm}/\gamma_\nu
-2m b\hspace{-2mm}/\gamma_\nu k\hspace{-2mm}/}{2(p\cdot k)^2}\right]u(p).
\label{eqa8}
\end{eqnarray}
The $b^2$ term is quite complicated. Making using of Eq.\,(17), we obtain
\begin{eqnarray}
M_\nu (b^2)&=&\bar{u}(q)\left[\gamma_\nu
\frac{1}{k\hspace{-2mm}/+q\hspace{-2mm}/-m}b\hspace{-2mm}/\gamma_5
\frac{1}{k\hspace{-2mm}/+q\hspace{-2mm}/-m}b\hspace{-2mm}/\gamma_5
\frac{1}{k\hspace{-2mm}/+q\hspace{-2mm}/-m}\gamma_0\right.
\nonumber\\
&&\left.+\gamma_0 
\frac{1}{p\hspace{-2mm}/-k\hspace{-2mm}/-m}b\hspace{-2mm}/\gamma_5
\frac{1}{p\hspace{-2mm}/-k\hspace{-2mm}/-m}b\hspace{-2mm}/\gamma_5
\frac{1}{p\hspace{-2mm}/-k\hspace{-2mm}/-m}\gamma_\nu\right]u(p)
\nonumber\\
&=&\bar{u}(q)\left[\frac{\gamma_\nu
\left(k\hspace{-2mm}/+q\hspace{-2mm}/+m\right)b\hspace{-2mm}/
\left(k\hspace{-2mm}/+q\hspace{-2mm}/+m\right)b\hspace{-2mm}/
\left(k\hspace{-2mm}/+q\hspace{-2mm}/+m\right)\gamma_0}{8(k\cdot q)^3}
\right.\nonumber\\
&&-\frac{\gamma_0
\left(p\hspace{-2mm}/-k\hspace{-2mm}/+m\right)b\hspace{-2mm}/
\left(p\hspace{-2mm}/-k\hspace{-2mm}/+m\right)b\hspace{-2mm}/
\left(p\hspace{-2mm}/-k\hspace{-2mm}/+m\right)\gamma_\nu}{8(k\cdot p)^3}
\nonumber\\
&&\left.-\frac{mb^2\gamma_\nu \left(k\hspace{-2mm}/+q\hspace{-2mm}/
+m\right)^2\gamma_0}{4(k\cdot q)^3}
+\frac{mb^2\gamma_0 \left(p\hspace{-2mm}/-k\hspace{-2mm}/
+m\right)^2\gamma_\nu}{4(k\cdot p)^3}\right]u(p)\nonumber\\
&\stackrel{k\rightarrow 0}{\simeq}&\bar{u}(q)
\left\{\frac{\left[
4 (k\cdot b+q\cdot b)^2-2k\cdot q b^2\right](2q_\nu-k\hspace{-2mm}/\gamma_\nu )
-4 (k+q)\cdot bk\cdot q\gamma_\nu
b\hspace{-2mm}/}{8(k\cdot q)^3}
\gamma_0\right.\nonumber\\
&&- \gamma_0\frac{\left[
4 (p\cdot b-k\cdot b)^2+2k\cdot p b^2\right]
(2p_\nu + \gamma_\nu  k\hspace{-2mm}/)
+4 (p-k)\cdot b k\cdot p b\hspace{-2mm}/\gamma_\nu}
{8(k\cdot p)^3} \nonumber\\
&&-\frac{mb^2(\gamma_\nu k\cdot q+2m q_\nu-mk\hspace{-2mm}/\gamma_\nu)
\gamma_0}{2(k\cdot q)^3}\nonumber\\
&&\left.+\frac{\gamma_0 mb^2(-k\cdot p\gamma_\nu
+2m p_\nu+m\gamma_\nu k\hspace{-2mm}/)}{2(k\cdot p)^3}\right\}u(p).
\label{eqa9}
\end{eqnarray}
In deriving Eqs.\,(\ref{eqa8}) and (\ref{eqa9}), we only keep
the IR singular terms as $k_\mu{\rightarrow}0$ and throw away the terms
containing $k_\nu$ and $k^2$ since $\epsilon^{(\lambda)}\cdot k=0$
and  $k^2=0$ on-shell.

 Eqs.\,(\ref{eqa6}) --- (\ref{eqa8}) give the amplitude to 
the second order of $b$,
\begin{eqnarray}
\epsilon_{\nu}^{(\lambda)}M^\nu_{(1\gamma)} &=& \epsilon^{(\lambda)}(k)\cdot 
\left[M (b^0)+M (b^1)+M (b^2)\right]\nonumber\\
&=& \bar{u}(q)\left[\gamma_0  M
+\gamma_0\gamma_5 M_5
+\gamma_\mu \gamma_0\gamma_5 M^{(1)}_{5\mu} 
+\gamma_5\gamma_0 \gamma_\mu M^{(2)}_{5\mu}
+\gamma_\mu \gamma_\nu \gamma_0\gamma_5 M^{(1)}_{5\mu\nu}
\right.\nonumber\\
&&+\gamma_5\gamma_0 \gamma_\mu \gamma_\nu M^{(2)}_{5\mu\nu}
+\gamma_\mu \gamma_\nu \gamma_\rho 
\gamma_0\gamma_5  M^{(1)}_{5\mu\nu\rho}
+\gamma_5\gamma_0 \gamma_\mu \gamma_\nu \gamma_\rho M^{(2)}_{5\mu\nu\rho}
+\gamma_\mu \gamma_0 M^{(1)}_{\mu}
\nonumber\\&& 
\left.+\gamma_0 \gamma_\mu M^{(2)}_{\mu}
+\gamma_\mu\gamma_\nu \gamma_0 M^{(1)}_{\mu\nu}
+\gamma_0\gamma_\nu \gamma_\mu M^{(2)}_{\mu\nu}\right]u(p),
\label{eq13a}
\end{eqnarray}
where the various $M$'s are listed as following,
\begin{eqnarray}
M&=&\frac{q\cdot\epsilon^{\lambda}}{q\cdot k}-
\frac{p\cdot\epsilon^{(\lambda)}}{p\cdot k}
+\frac{q\cdot\epsilon^{(\lambda)}
\left[2 (k\cdot b+q\cdot b)^2-k\cdot qb^2-2 m^2b^2\right]}{2(k\cdot q)^3}
\nonumber\\
&&-\frac{p\cdot\epsilon^{(\lambda)}
\left[2 (p\cdot b-k\cdot b)^2+k\cdot pb^2-2 m^2b^2\right]}{2(k\cdot p)^3};
\nonumber \\
M_5&=&\frac{q\cdot\epsilon^{(\lambda)}(k+q)\cdot b}{(q\cdot k)^2}-
\frac{p\cdot\epsilon^{(\lambda)}(p-k)\cdot b}{(p\cdot k)^2};\nonumber  \\
M^{(1)}_{5\mu}&=&-\frac{q\cdot\epsilon^{(\lambda)}m b_\mu}{(q\cdot k)^2};~~~~
M^{(2)}_{5\mu}=-\frac{p\cdot\epsilon^{(\lambda)}m b_\mu}{(p\cdot k)^2};
\nonumber \\
M^{(1)}_{5\mu\nu}&=&-\frac{q\cdot b k_\mu \epsilon^{(\lambda)}_\nu
+k\cdot q \epsilon^{(\lambda)}_\mu b_\nu}{2 (q\cdot k)^2};~~~~
M^{(2)}_{5\mu\nu}=\frac{p\cdot b \epsilon^{(\lambda)}_\mu k_\nu
+k\cdot p b_\mu \epsilon^{(\lambda)}_\nu}{2 (q\cdot k)^2};\nonumber  \\
M^{(1)}_{5\mu\nu\rho}&=&\frac{mk_\mu \epsilon^{(\lambda)}_\nu b_\rho}
{2 (q\cdot k)^2}; ~~~~
M^{(2)}_{5\mu\nu\rho}=\frac{mb_\mu \epsilon^{(\lambda)}_\nu k_\rho}
{2 (p\cdot k)^2};\nonumber  \\
M^{(1)}_{\mu}&=&-\frac{mb^2q\cdot k \epsilon^{(\lambda)}}
{2(q\cdot k)^3};~~~~
M^{(2)}_{\mu}=-\frac{mb^2p\cdot k\epsilon^{(\lambda)}}{2(p\cdot k)^3};
\nonumber \\
M^{(1)}_{\mu\nu}&=& 
\frac{[(k\cdot q +2 m^2)b^2-2 (k\cdot b+q\cdot b)^2]
 k_\mu \epsilon^{(\lambda)}_\nu 
-2(k+q)\cdot b k\cdot q \epsilon^{(\lambda)}_\mu b_\nu}
{4(k\cdot q)^3};\nonumber  \\
M^{(2)}_{\mu\nu}&=&\frac{[(-k\cdot p +2 m^2)b^2-2 (p\cdot b-k\cdot b)^2]
 \epsilon^{(\lambda)}_\mu k_\nu
-2(p-k)\cdot b k\cdot p b_\mu \epsilon^{(\lambda)}_\nu }
{4(k\cdot p)^3}.
\label{eq14a}
\end{eqnarray}
Using the following well-known formula valid for a general operator 
$\Gamma$,
\begin{eqnarray}
|\bar{u}(q)\Gamma u(p)|^2=\bar{u}(q)\Gamma u(p)\bar{u}(p)\gamma_0
\Gamma^{\dagger}\gamma_0 u(q)
=\mbox{Tr}\left[\frac{q\hspace{-2mm}/+m}{2m}\Gamma\frac{p\hspace{-2mm}/+m}{2m}
\gamma_0\Gamma^{\dagger}\gamma_0\right],
\label{eq15a}
\end{eqnarray}
we can evaluate $\sum_\lambda |\epsilon^{(\lambda)}\cdot M|^2$. 
For instance,
\begin{eqnarray}
|\bar{u}(q)\gamma_0 u(p)M|^2
&=&\frac{1}{m^2}\left(2E_p E_q-p\cdot q+m^2\right) |M|^2;\nonumber  \\ 
|\bar{u}(q)\gamma_0\gamma_5 u(p)M|^2
&=&\frac{1}{m^2}\left(2E_p E_q-p\cdot q-m^2\right) |M_5|^2;\nonumber  \\ 
|\bar{u}(q)\gamma_\mu\gamma_\nu\gamma_0 u(p)M^{\mu\nu}|^2
&=&\frac{1}{4m^2}\mbox{Tr}\left[\left(q\hspace{-2mm}/+m\right)
\gamma_{\mu}\gamma_{\nu}\gamma_0 \left(p\hspace{-2mm}/+m\right)
\gamma_0\gamma_\rho\gamma_{\lambda}\right]M^{\mu\nu}M^{*\lambda\rho}
\nonumber\\
&=&\frac{1}{m^2}\left\{2E_p\left[q_\mu \left(g_{\nu 0}g_{\rho\lambda}
-g_{\nu\rho}g_{0\lambda}+g_{\nu\lambda}g_{0\rho}\right)
-\left(\mu\longleftrightarrow\nu\right)\right]\right.\nonumber\\
&&-2E_p\left[q_\rho \left(g_{\mu\nu}g_{0\lambda}-g_{\mu 0}g_{\nu\lambda}
+g_{\mu\lambda}g_{\nu 0}\right)-\left(\lambda\longleftrightarrow\rho\right)
\right]\nonumber\\
&&-\left[q_{\mu}\left(p_\nu g_{\rho\lambda}-g_{\nu\rho}p_{\lambda}-
g_{\nu\lambda}p_{\rho}\right)-\left(\mu\longleftrightarrow\nu\right)\right]
\nonumber\\
&&-\left[q_{\rho}\left(g_{\mu\nu}p_\lambda -g_{\nu\lambda}p_{\mu}-
g_{\mu\lambda}p_{\nu}\right)+\left(\lambda\longleftrightarrow\rho\right)\right]
\nonumber\\
&+&\left.\left(2E_pE_q-p\cdot q+m^2\right)\left(g_{\mu\nu}g_{\rho\lambda}
-g_{\mu\rho}g_{\nu\lambda}+g_{\mu\lambda}g_{\nu\rho}\right)\right\}
M^{\mu\nu}M^{*\lambda\rho};
\label{eq16a}
\end{eqnarray}
There are 132 terms in the expansion of the $|\epsilon\cdot M|^2$, and
some of them may vanish . 
In principle, with this expansion, using the fact
that $|b|{\ll}m$ and taking
the non-relativistic limit,
$|{\bf p}|, |{\bf q}|{\ll}m$,
\begin{eqnarray}
p{\cdot}k=E_qE_k-{\bf p}\cdot{\bf k}=m\sqrt{1+{\bf p}^2/m^2} E_k
-{\bf p}\cdot{\bf k}{\simeq}m E_k-{\bf p}\cdot{\bf k}, 
\label{eq17a}
\end{eqnarray}
we can calculate the contribution from photons with long
wave length and observe whether the IR divergence associated with 
the soft photon emission cancels the divergent term in
the vertex correction\cite{gre}. However, Eqs.\,(\ref{eq13a}) -- (\ref{eq17a})
imply that due to the various terms relevant to $b_\mu$ in (\ref{eq14a}),
the following relation in the soft photon approximation cannot be valid
as in the conventional QED,
\begin{eqnarray}
\sum_{\lambda=1,2}|\epsilon_{\nu}^{(\lambda)}(k)M_{(1\gamma)}^\nu|^2|
=e^2|M^{(0)}|^2 f(k\cdot p, k\cdot q, l^2).
\label{eq20a}
\end{eqnarray}
As shown below, this relation plays a key role 
in the cancellation of IR divergences in conventional QED. Thus
the non-existence of (\ref{eq20a}) further enforces the difficulty
in implementing the IR divergence cancellation.

To explicitly see whether or not the IR divergences cancel,  we 
define the physical ``measurable'' cross section as in conventional 
QED\cite{ref15},
\begin{eqnarray}
\sigma =\int_0^{\Delta E}d{\cal E}\frac{d\left(\sigma_1
 +\sigma_{1\gamma}\right)}{d{\cal E}},
\label{eq27a}
\end{eqnarray}
where $\Delta E$ is the energy resolution
of the detection device and ${\cal E}=E_q-E_p$ is the energy of
the emitted real photon. In standard QED, Eq.\,(\ref{eq20a})  leads
to 
\begin{eqnarray}
\int_0^{\Delta E}d{\cal E}\frac{d \sigma_{1\gamma}}{d{\cal E}}
\stackrel{k_\mu\rightarrow 0}{\simeq}
\int_0^{\Delta E}d{\cal E}\frac{d \sigma_0}{d{\cal E}}
\int_0^{|{\bf k}|=E_k{\leq}\Delta E}\frac{d^3{\bf k}}{E_k(2\pi)^3}e^2|M_0|^2,
\end{eqnarray} 
where the integration over $d^3{\bf k}$ can be performed with
dimension regularization\cite{ref15}, 
i.e. taking $3\longrightarrow n-1=3+\epsilon_{IR}$. 
Consequently, the differential cross section of an electron interacting
with an external Coulomb potential in  conventional QED has the following
simple form:
\begin{eqnarray}
d\sigma =d\sigma_0\left[1+2 F_1(l^2)+\int_0^{\Delta E}
\frac{d^{3+\epsilon_{IR}}{\bf k}}{(2\pi)^3|{\bf k}|}e^2|M_0|^2\right].
\label{eq23a}
\end{eqnarray} 
with no remaining IR divergences.

Unfortunately,
the relations (\ref{eq5a}), (\ref{eq6a}) and (\ref{eq20a}) do not
exist in QED with Lorentz and $CPT$ violation term,  due to the complicated 
tensor structure of
the matrix elements listed in Eqs.\,(\ref{eq8a26})
and (\ref{eq13a}). This makes it impossible for the physical
cross section to reduce to the form of (\ref{eq23a}). 
In fact, even if the relation (\ref{eq23a}) could be established,
this would not guarantee the cancellation of the IR divergence. The first 
direct reason is that
the form factors $F_i$ with $i{\geq}2$ also contain  
IR divergences induced by the $\bar{\psi}b\hspace{-2mm}/\gamma_5\psi$
 term; The second one is more catastrophic: the
bremsstrahlung process in the soft photon limit contains
some novel IR divergences. For example, we find the following term
in evaluating $|\epsilon^{(\lambda)}\cdot M_{(1\gamma)}|^2$, 
\begin{eqnarray}
\bar{u}(q)\gamma_0 u(p)M\left[\bar{u}(q)\gamma_\mu\gamma_\nu\gamma_0
\gamma_5 u(p)M^{(1)\mu\nu}_5\right]^{\dagger}
=-\frac{i}{m^2}\left(\epsilon_{\mu\nu\lambda\rho}p^{\lambda}q^{\rho}
+2E_p\epsilon_{0\mu\nu\rho}q^{\rho}\right)M M^{*\mu\nu}.
\end{eqnarray} 
It is clear that there can be no such IR divergent term
in the vertex correction. This fact can be verified  by simply
 comparing the various tensor structures in Eqs.\,(\ref{eq8a26}) 
and (\ref{eq13a}).

The above discussion has shown that the introduction
of the Lorentz and $CPT$ violating term, 
$\bar{\psi}b\hspace{-2mm}/\gamma_5\psi$ in the fermionic sector of
QED gives rise to   IR divergences in  the on-shell vertex radiative correction
that cannot be cured by considering the soft emission of bremsstrahlung. 
As a consequence, the theoretical value of 
the Lamb shift inevitably becomes unphysical.
This suggests that it may not be appropriate to
investigate  Lorentz and 
$CPT$ violation effects in the electromagnetic
interaction by simply modifying  the fermionic
sector of QED.

\section{Summary and Discussion}

We have calculated the one-loop polarization tensor and the on-shell vertex 
radiative correction to  second order in $b$ for QED with an additional
$CPT$-odd term $\psi b\hspace{-2mm}/\gamma_5\psi$ in the fermionic sector. 
This term is responsible for the generation of Lorentz-$CPT$ violation through
radiative corrections. Furthermore, we showed explicitly the resulting 
$b_\mu$-dependence of the electron anomalous magnetic moment 
and the Lamb shift, demonstrating that the Lorentz and $CPT$ violation 
term in the fermionic sector gives rise to remarkable effects 
on these two important physical predictions of QED. However, both 
expressions for the anomalous magnetic moment and the Lamb shift 
were shown to contain IR divergences linked to the Lorentz and $CPT$ 
violation. The IR divergent terms in the anomalous magnetic
moment lead to non-physical effective interaction and the IR divergence 
in the Lamb shift cannot be canceled in physical 
cross-sections by the contribution from the bremsstrahlung. 
This seems to  imply that the Lorentz and $CPT$ violation term
must vanish. Of course, our result does not negate the possible existence
of Lorentz and $CPT$  violation phenomena in the electromagnetic interaction
in general. It only means that it may not be 
appropriate to explore theoretically
the Lorentz and $CPT$ violating effects by putting explicit
violation terms in the fermionic sector. Thus alternative 
models may be required.  Our main aim here
is to reveal the possible effects of a 
$CPT$-odd term on the anomalous magnetic moment
and Lamb shift data, with  emphasis on their field theoretic origins 
and to provide constraints on theoretical models that may be
used to explore Lorentz and $CPT$ violation in  electromagnetic
phenomena.

Finally, we note that
 the $b$-dependent part of the vertex radiative correction 
leads to many new types of non-minimal coupling between the electron and 
photon at the second order of $b$ such as 
$(b\cdot \partial)^2\bar{\psi}A\hspace{-2mm}/\psi$, 
$\bar{\psi}A\hspace{-2mm}/ (b\cdot \partial)^2\psi$, 
$(b\cdot\partial)\bar{\psi}A\hspace{-2mm}/(b\cdot\partial)\psi$
and $\bar{\psi}b\hspace{-2mm}/\psi b\cdot A$ etc. These non-minimal
interactions  can  yield even more remarkable effects than the anomalous
magnetic moment and the Lamb shift at low-energy, but they are difficult
to calculate explicitly. For example, to first order in $b$ there arise
couplings of the form $(b\cdot \partial)\bar{\psi}A\hspace{-2mm}/\gamma_5\psi$,
$\bar{\psi}A\hspace{-2mm}/\gamma_5(b\cdot \partial)\psi$,
$\partial_\mu\bar{\psi}b\hspace{-2mm}/\gamma_5\psi  A^\mu$
and $\bar{\psi}b\hspace{-2mm}/\gamma_5\partial_\mu\psi  A^\mu$ etc.
These are  non-minimal couplings between
vector field and axial vector currents and hence are not explicitly
invariant under electric charge conjugation. They will
lead to  electrostatic interactions in which particles with the same charge
attract whereas  opposite charges repel.

\acknowledgements

This work is supported in part by the Natural Sciences and Engineering
Research Council of Canada. We would like to thank Professor R. Jackiw
for drawing our attention to this topic.

\begin{figure}
\centering
\input FEYNMAN

\begin{picture}(40000,18000)
\THICKLINES
\drawline\fermion[\SW\REG](12000,10000)[2000]
\drawline\photon[\W\REG](\pbackx,\pbacky)[3]
\drawline\photon[\S\REG](\photonbackx,\photonbacky)[3]
\drawline\fermion[\SW\REG](\photonbackx,\photonbacky)[3000]
\drawline\fermion[\NE\REG](\pfrontx,\pfronty)[6000]
\drawline\photon[\N\REG](\pbackx,\pbacky)[5]
\drawline\fermion[\SE\REG](\photonfrontx,\photonfronty)[9100]
\drawarrow[\NE\ATBASE](6700,4700)
\drawarrow[\SE\ATBASE](17000,4700)
\put(4500,4000){$p$}
\put(18500,4000){$q$}

\drawline\fermion[\SE\REG](30000,10000)[2000]
\drawline\photon[\E\REG](\pbackx,\pbacky)[3]
\drawline\photon[\S\REG](\photonbackx,\photonbacky)[3]
\drawline\fermion[\SE\REG](\photonbackx,\photonbacky)[3000]
\drawline\fermion[\NW\REG](\pfrontx,\pfronty)[6000]
\drawline\photon[\N\REG](\pbackx,\pbacky)[5]
\drawline\fermion[\SW\REG](\photonfrontx,\photonfronty)[9100]
\drawarrow[\NE\ATBASE](25000,4700)
\drawarrow[\SE\ATBASE](35300,4700)
\put(22000,4000){$p$}
\put(37000,4000){$q$}
\end{picture}

\begin{picture}(40000,16000)
\THICKLINES
\drawline\photon[\S\REG](12000,15000)[5]
\drawline\fermion[\SW\REG](\photonbackx,\photonbacky)[9000]
\drawline\fermion[\SE\REG](\photonbackx,\photonbacky)[9000]
\drawarrow[\NE\ATBASE](6700,4700)
\drawarrow[\SE\ATBASE](17300,4700)
\put(4000,4000){$p$}
\put(19000,4000){$q$}
\put(9300,7500){$\times$}
\put(21000,200){$(a)$}

\drawline\photon[\S\REG](30000,15000)[5]
\drawline\fermion[\SW\REG](\photonbackx,\photonbacky)[9000]
\drawline\fermion[\SE\REG](\photonbackx,\photonbacky)[9000]
\drawarrow[\NE\ATBASE](24700,4700)
\drawarrow[\SE\ATBASE](35300,4700)
\put(22000,4000){$p$}
\put(37000,4000){$q$}
\put(31800,7500){$\times$}
\end{picture}

\begin{picture}(40000,16000)
\THICKLINES
\drawline\photon[\S\REG](10000,15000)[5]
\drawline\fermion[\SW\REG](\photonbackx,\photonbacky)[6400]
\drawline\fermion[\W\REG](\fermionbackx,\fermionbacky)[4000]
\drawarrow[\E\ATBASE](\pmidx,\pmidy)
\put(3000,6000){$p$}
\drawline\fermion[\SE\REG](\photonbackx,\photonbacky)[6400]
\drawline\fermion[\E\REG](\fermionbackx,\fermionbacky)[4000]
\drawarrow[\E\ATBASE](\pmidx,\pmidy)
\put(16000,6000){$q$}
\drawline\photon[\W\REG](\fermionfrontx,\fermionfronty)[9]
\put(10000,1500){$(b)$}

\drawline\photon[\S\REG](30000,14300)[3]
\put(30000,10000){\circle{3000}}
\drawline\photon[\S\REG](30000,8600)[2]
\drawline\fermion[\SW\REG](\photonbackx,\photonbacky)[4000]
\drawarrow[\NE\ATBASE](\pmidx,\pmidy)
\drawline\fermion[\SE\REG](\photonbackx,\photonbacky)[4000]
\drawarrow[\SE\ATBASE](\pmidx,\pmidy)
\put(26000,4500){$p$}
\put(33000,4500){$q$}
\put(30000,1500){$(c)$}
\end{picture}

\caption{\protect\small  One-loop Feynman diagrams contributing to Lamb
shift: (a) the electron self-energy contribution, which actually vanishes
in mass-shell renormalization scheme, $\times$ representing the counterterm
for electron mass renormalization; (b) the vertex radiative correction; (c)
the contribution from vacuum polarization tensor.}

\begin{picture}(50000,8000)
\THICKLINES
\put(20000,5000){\circle{3000}}
\startphantom
\drawline\photon[\E\REG](0,0)[2]
\stopphantom
\pbackx=18400 \pbacky=5000
\global\multiply\plengthx by -1
\global\multiply\plengthy by -1
\global\advance\pbackx by \plengthx
\global\advance\pbacky by \plengthy
\drawline\photon[\E\REG](\pbackx,\pbacky)[2]
\drawline\photon[\W\FLIPPED](\photonfrontx,\photonfronty)[2]
\photonbackx=21400 \photonbacky=5000
\negate\photonlengthx
\negate\photonlengthy
\global\advance\photonbackx by \photonlengthx
\global\advance\photonbacky by \photonlengthy
\drawline\photon[\W\FLIPPED](\photonbackx,\photonbacky)[2]
\drawline\photon[\E\REG](\photonfrontx,\photonfronty)[2]
\put(20000,1500){$(a)$}
\end{picture}

\begin{picture}(50000,8000)
\THICKLINES
\put(13500,6080){$\otimes$}
\put(14000,5000){\circle{3000}}
\startphantom
\drawline\photon[\E\REG](0,0)[2]
\stopphantom
\pbackx=12400 \pbacky=5000
\global\multiply\plengthx by -1
\global\multiply\plengthy by -1
\global\advance\pbackx by \plengthx
\global\advance\pbacky by \plengthy
\drawline\photon[\E\REG](\pbackx,\pbacky)[2]
\drawline\photon[\W\FLIPPED](\photonfrontx,\photonfronty)[2]
\photonbackx=15400 \photonbacky=5000
\negate\photonlengthx
\negate\photonlengthy
\global\advance\photonbackx by \photonlengthx
\global\advance\photonbacky by \photonlengthy
\drawline\photon[\W\FLIPPED](\photonbackx,\photonbacky)[2]
\drawline\photon[\E\REG](\photonfrontx,\photonfronty)[2]

\THICKLINES
\put(28500,3280){$\otimes$}
\put(29000,5000){\circle{3000}}
\startphantom
\drawline\photon[\E\REG](0,0)[2]
\stopphantom
\pbackx=27400 \pbacky=5000
\global\multiply\plengthx by -1
\global\multiply\plengthy by -1
\global\advance\pbackx by \plengthx
\global\advance\pbacky by \plengthy
\drawline\photon[\E\REG](\pbackx,\pbacky)[2]
\drawline\photon[\W\FLIPPED](\photonfrontx,\photonfronty)[2]
\photonbackx=30400 \photonbacky=5000
\negate\photonlengthx
\negate\photonlengthy
\global\advance\photonbackx by \photonlengthx
\global\advance\photonbacky by \photonlengthy
\drawline\photon[\W\FLIPPED](\photonbackx,\photonbacky)[2]
\drawline\photon[\E\REG](\photonfrontx,\photonfronty)[2]
\put(20000,1500){$(b)$} 
\end{picture}

\begin{picture}(50000,8000)
\THICKLINES
\put(7500,6080){$\otimes$}
\put(8000,5000){\circle{3000}}
\put(7500,3280){$\otimes$}
\startphantom
\drawline\photon[\E\REG](0,0)[2]
\stopphantom
\pbackx=6400 \pbacky=5000
\global\multiply\plengthx by -1
\global\multiply\plengthy by -1
\global\advance\pbackx by \plengthx
\global\advance\pbacky by \plengthy
\drawline\photon[\E\REG](\pbackx,\pbacky)[2]
\drawline\photon[\W\FLIPPED](\photonfrontx,\photonfronty)[2]
\photonbackx=9400 \photonbacky=5000
\negate\photonlengthx
\negate\photonlengthy
\global\advance\photonbackx by \photonlengthx
\global\advance\photonbacky by \photonlengthy
\drawline\photon[\W\FLIPPED](\photonbackx,\photonbacky)[2]
\drawline\photon[\E\REG](\photonfrontx,\photonfronty)[2]

\THICKLINES
\put(20600,5700){$\otimes$}
\put(22000,5000){\circle{3000}}
\put(22600,5700){$\otimes$}
\startphantom
\drawline\photon[\E\REG](0,0)[2]
\stopphantom
\pbackx=20400 \pbacky=5000
\global\multiply\plengthx by -1
\global\multiply\plengthy by -1
\global\advance\pbackx by \plengthx
\global\advance\pbacky by \plengthy
\drawline\photon[\E\REG](\pbackx,\pbacky)[2]
\drawline\photon[\W\FLIPPED](\photonfrontx,\photonfronty)[2]
\photonbackx=23400 \photonbacky=5000
\negate\photonlengthx
\negate\photonlengthy
\global\advance\photonbackx by \photonlengthx
\global\advance\photonbacky by \photonlengthy
\drawline\photon[\W\FLIPPED](\photonbackx,\photonbacky)[2]
\drawline\photon[\E\REG](\photonfrontx,\photonfronty)[2]

\THICKLINES

\put(35600,3600){$\otimes$}
\put(37000,5000){\circle{3000}}
\put(37400,3600){$\otimes$}
\startphantom
\drawline\photon[\E\REG](0,0)[2]
\stopphantom
\pbackx=35400 \pbacky=5000
\global\multiply\plengthx by -1
\global\multiply\plengthy by -1
\global\advance\pbackx by \plengthx
\global\advance\pbacky by \plengthy
\drawline\photon[\E\REG](\pbackx,\pbacky)[2]
\drawline\photon[\W\FLIPPED](\photonfrontx,\photonfronty)[2]
\photonbackx=38400 \photonbacky=5000
\negate\photonlengthx
\negate\photonlengthy
\global\advance\photonbackx by \photonlengthx
\global\advance\photonbacky by \photonlengthy
\drawline\photon[\W\FLIPPED](\photonbackx,\photonbacky)[2]
\drawline\photon[\E\REG](\photonfrontx,\photonfronty)[2]
\put(20000,1500){$(c)$} 
\end{picture}

\caption{\protect\small Vacuum polarization up to the second order
of $b_\mu$ contributed by fermionic loops with various insertions 
of $CPT$-odd vertex $b\hspace{-1.6mm}/\gamma_5$ in the internal
fermionic lines, $\otimes$ denoting the vertex 
$b\hspace{-1.8mm}/\gamma_5$.}

\begin{picture}(40000,18000)
\THICKLINES
\drawline\photon[\S\REG](20000,15000)[5]
\drawline\fermion[\SW\REG](\photonbackx,\photonbacky)[6400]
\drawline\fermion[\W\REG](\fermionbackx,\fermionbacky)[4000]
\drawarrow[\E\ATBASE](\pmidx,\pmidy)
\put(13000,6000){$p$}
\drawline\fermion[\SE\REG](\photonbackx,\photonbacky)[6400]
\drawline\fermion[\E\REG](\fermionbackx,\fermionbacky)[4000]
\drawarrow[\E\ATBASE](\pmidx,\pmidy)
\put(26000,6000){$q$}
\drawline\photon[\W\REG](\fermionfrontx,\fermionfronty)[9]
\put(19000,2500){$(a)$}
\end{picture}

\begin{picture}(40000,16000)
\THICKLINES
\drawline\photon[\S\REG](10000,15000)[5]
\drawline\fermion[\SW\REG](\photonbackx,\photonbacky)[6400]
\drawline\fermion[\W\REG](\fermionbackx,\fermionbacky)[4000]
\drawarrow[\E\ATBASE](\pmidx,\pmidy)
\put(3000,6000){$p$}
\drawline\fermion[\SE\REG](\photonbackx,\photonbacky)[6400]
\drawline\fermion[\E\REG](\fermionbackx,\fermionbacky)[4000]
\drawarrow[\E\ATBASE](\pmidx,\pmidy)
\put(16000,6000){$q$}
\drawline\photon[\W\REG](\fermionfrontx,\fermionfronty)[9]
\put(7300,7500){$\otimes$}
\put(20000,2500){$(b)$}

\drawline\photon[\S\REG](30000,15000)[5]
\drawline\fermion[\SW\REG](\photonbackx,\photonbacky)[6400]
\drawline\fermion[\W\REG](\fermionbackx,\fermionbacky)[4000]
\drawarrow[\E\ATBASE](\pmidx,\pmidy)
\put(23000,6000){$p$}
\drawline\fermion[\SE\REG](\photonbackx,\photonbacky)[6400]
\drawline\fermion[\E\REG](\fermionbackx,\fermionbacky)[4000]
\drawarrow[\E\ATBASE](\pmidx,\pmidy)
\put(36000,6000){$q$}
\drawline\photon[\W\REG](\fermionfrontx,\fermionfronty)[9]
\put(31800,7500){$\otimes$}
\end{picture}

\begin{picture}(40000,16000)
\THICKLINES
\drawline\photon[\S\REG](5000,15000)[4]
\drawline\fermion[\SW\REG](\photonbackx,\photonbacky)[5400]
\drawline\fermion[\W\REG](\fermionbackx,\fermionbacky)[3000]
\drawarrow[\E\ATBASE](\pmidx,\pmidy)
\put(50,6000){$p$}
\drawline\fermion[\SE\REG](\photonbackx,\photonbacky)[5400]
\drawline\fermion[\E\REG](\fermionbackx,\fermionbacky)[3000]
\drawarrow[\E\ATBASE](\pmidx,\pmidy)
\put(9000,6000){$q$}
\drawline\photon[\W\REG](\fermionfrontx,\fermionfronty)[8]
\put(3500,9700){$\otimes$}
\put(2500,8700){$\otimes$}

\drawline\photon[\S\REG](20000,15000)[4]
\drawline\fermion[\SW\REG](\photonbackx,\photonbacky)[5400]
\drawline\fermion[\W\REG](\fermionbackx,\fermionbacky)[3000]
\drawarrow[\E\ATBASE](\pmidx,\pmidy)
\put(15000,6000){$p$}
\drawline\fermion[\SE\REG](\photonbackx,\photonbacky)[5400]
\drawline\fermion[\E\REG](\fermionbackx,\fermionbacky)[3000]
\drawarrow[\E\ATBASE](\pmidx,\pmidy)
\put(24000,6000){$q$}
\drawline\photon[\W\REG](\fermionfrontx,\fermionfronty)[8]
\put(20600,9700){$\otimes$}
\put(21600,8700){$\otimes$}
\put(20000,2500){$(c)$}

\drawline\photon[\S\REG](35000,15000)[4]
\drawline\fermion[\SW\REG](\photonbackx,\photonbacky)[5400]
\drawline\fermion[\W\REG](\fermionbackx,\fermionbacky)[3000]
\drawarrow[\E\ATBASE](\pmidx,\pmidy)
\put(30000,6000){$p$}
\drawline\fermion[\SE\REG](\photonbackx,\photonbacky)[5400]
\drawline\fermion[\E\REG](\fermionbackx,\fermionbacky)[3000]
\drawarrow[\E\ATBASE](\pmidx,\pmidy)
\put(40000,6000){$q$}
\drawline\photon[\W\REG](\fermionfrontx,\fermionfronty)[8]
\put(36000,9200){$\otimes$}
\put(33000,9200){$\otimes$}
\end{picture}

\caption{\protect\small One-loop vertex correction up to the second order
of $b_\mu$ with various insertions
of $CPT$-odd vertex $b\hspace{-1.6mm}/\gamma_5$ in the internal
fermionic lines, $\otimes$ denoting the vertex
$b\hspace{-1.8mm}/\gamma_5$.}

\end{figure}

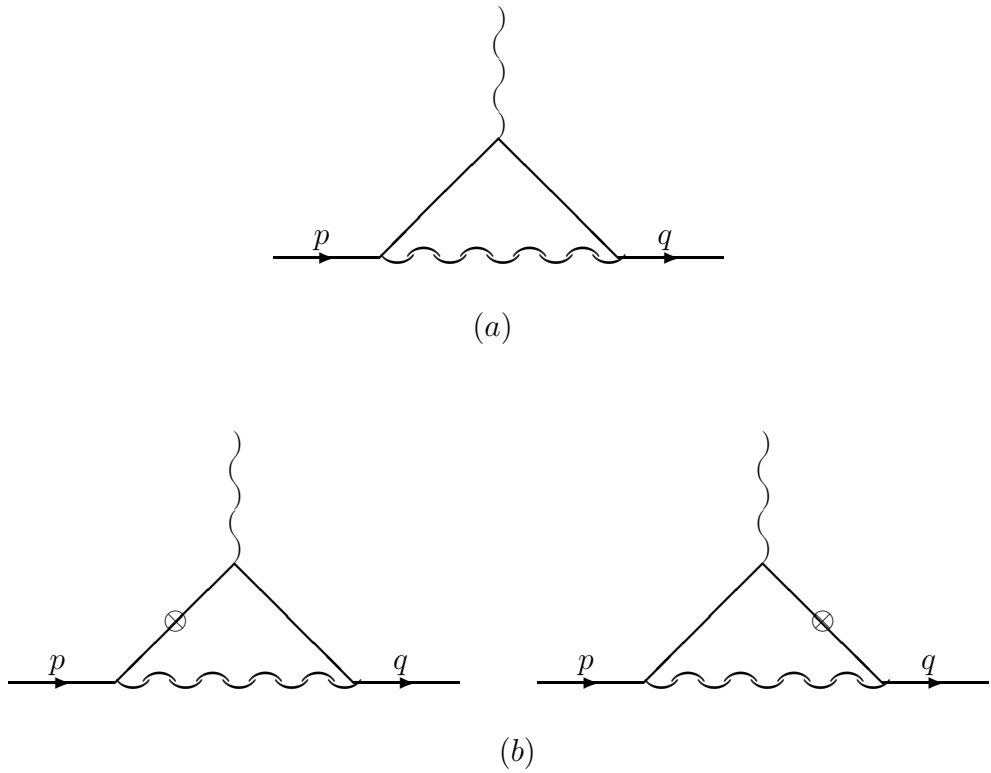
\end{document}